\documentclass[prx,superscriptaddress,floatfix,nofootinbib,a4paper,twocolumn,aps,10pt]{revtex4-1}
\usepackage[utf8]{inputenc}
\usepackage{natbib}
\usepackage{amsmath}
\usepackage{amsbsy}
\usepackage{amssymb}
\usepackage{scrextend}
\usepackage{graphicx}
\usepackage{color}
\usepackage[usenames, dvipsnames]{xcolor}
\usepackage[colorlinks=true, allcolors = Bittersweet]{hyperref}
\renewcommand{\vec}{\boldsymbol}
\newcommand{\kB}{k_{\scriptscriptstyle B}}

\newcommand{\D}{\mathfrak{D}}
\newcommand{\T}{\mathfrak{T}}

\DeclareMathOperator{\sign}{sign}
\DeclareMathOperator{\Erf}{Erf}
\DeclareMathOperator{\Erfc}{Erfc}
\DeclareMathOperator{\arccot}{arccot}
\DeclareMathOperator{\tr}{tr}
\DeclareMathOperator{\Poiss}{Poiss}
\begin{document}
\title{Hessian characterization of a vortex in a maze}
\author{R.\ Willa}
\affiliation{Institute for Theory of Condensed Matter, Karlsruhe Institute of
Technology, 76131 Karlsruhe, Germany}
\affiliation{Heidelberger Akademie der Wissenschaften, 69117 Heidelberg, Germany}

\author{V.B.\ Geshkenbein}
\affiliation{Institute for Theoretical Physics, ETH Zurich, 8093 Zurich,
Switzerland}
\author{G.\ Blatter}
\affiliation{Institute for Theoretical Physics, ETH Zurich, 8093 Zurich,
Switzerland}
\date{\today}

\begin{abstract}
Recent advances in vortex imaging allow for tracing the position of individual
vortices with high resolution. Pushing an isolated vortex through the sample
with the help of a controlled $dc$ transport current and measuring its local
$ac$ response, the pinning energy landscape could be reconstructed along the
vortex trajectory [L.\ Embon \textit{et al.}, Scientific Reports \textbf{5},
7598 (2015)]. This setup with linear tilts of the potential landscape reminds
about the dexterity game where a ball is balanced through a maze. The
controlled motion of objects through such tilted energy landscapes is
fundamentally limited to those areas of the landscape developing local minima
under appropriate tilt. We introduce the Hessian stability map and the Hessian
character of a pinning landscape as new quantities to characterize a pinning
landscape. We determine the Hessian character, the area fraction admitting
stable vortex positions, for various types of pinning potentials: assemblies
of cut parabolas,  Lorentzian- and Gaussian-shaped traps, as well as a
Gaussian random disordered energy landscape, with the latter providing a
universal result of $(3-\sqrt{3})/6 \approx 21\%$ of stable area. 
Furthermore, we discuss various aspects of the vortex-in-a-maze experiment.
\end{abstract}

\maketitle

\section{Introduction}
\begin{figure}[b]
\centering
\includegraphics[width = \linewidth]{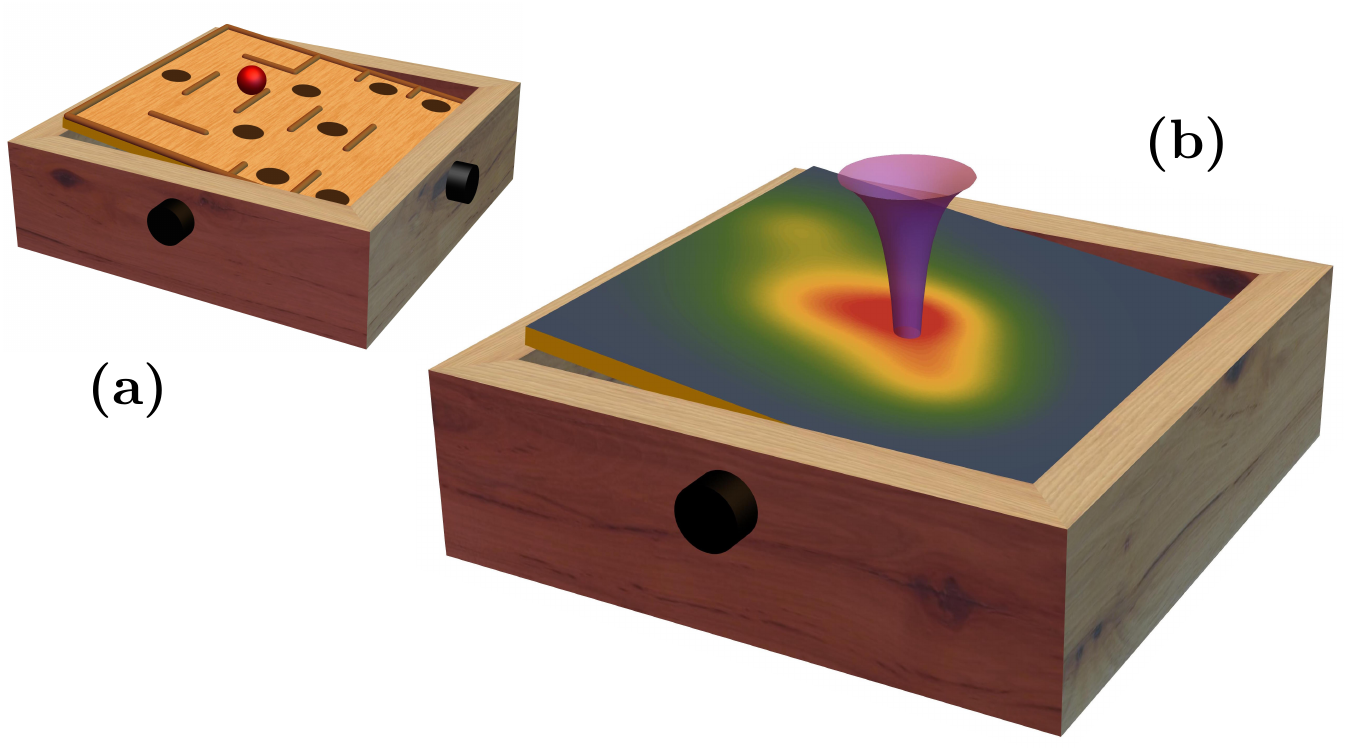}
\caption{(a) Ball in the maze: In this dexterity game, a ball driven by
gravity is guided through a labyrinth by adjusting the slope of the game board
via the two handles (front and right knobs). (b) Vortex in the maze: the color
map on the game-board shows the pinning potential landscape derived in Ref.~\cite{Embon2015}, see figure 5(a) therein. When compared to the
situation in (a), the labyrinth is replaced by the pinscape, while the
gravitational force manipulated by tilt is replaced by the Lorentz force
acting on the vortex; this corresponds to a tilt in only one direction as
indicated by the single knob (front). While this feature limits the region
that is probed by the vortex, combining low- and high-frequency response data
as well as different entry points in principle allows for extended vortex
guiding (and thus reconstruction of the two-dimensional pinscape) within the
stable regions of the potential, see Fig.~\ref{fig:stability-region}.}
\label{fig:mazes}
\end{figure}

The recent years have seen an astounding progress in the ability to image
vortices in superconductors \cite{Tonomura2001, Bending1999, Kirtley2010,
Suderow2014, Thiel2016}.  The high accuracy of these local-probe techniques
allow to study the shape of individual vortices \cite{Thiel2016} and even
manipulate them, e.g., via magnetic forces \cite{Straver2008,
Auslaender2009} or local mechanical stress \cite{Kremen2016}.
A new quality in precision-imaging has been achieved using a novel
SQUID-on-Tip (SOT) device combined with $ac$ techniques \cite{Embon2015, Embon2017}.
Changing the current drive in the sample allows to push and trace individual
vortices and extract the shape of the energy landscape (pinning landscape or
simply pinscape) from measured SOT data. Such information is most welcome in
optimizing pinscapes, which in turn is of great technological interest for
high-current applications \cite{Kwok2016, Sadovskyy2016b}. The functionality
of the experiment reminds about the well-known `ball-in-the-maze' dexterity
game shown in Fig.~\ref{fig:mazes}(a), where a ball is driven through a maze
by controlling the tilt of the plane. The present work focuses on the
`vortex-in-the-maze' problem, see Fig.~\ref{fig:mazes}(b), where a vortex is
driven across a pinning landscape through a controlled transport current that
induces a linear tilt of the potential. Here, we address the question which
parts of the pinning energy landscape can be probed in such an experiment,
that takes us to the \emph{Hessian stability map} as a new charateristics of a
pinscape.  The Hessian map of a pinning landscape then defines the areal
regions where vortices can assume stable positions---vortex trajectories
realizable in the vortex-in-a-maze setup are limited to these stable areas. We
define the \emph{Hessian character} of a landscape as the area fraction of the
plane where vortices can be pinned and determine this quantity for various
types of pinning landscapes, a random distribution of cut parabolic wells and
of Lorentzian- and Gaussian-shaped pins of given density; such traps are
often used in
numerical work \cite{Reichhardt1995, Olson2017} on vortex pinning and
dynamics.  Furthermore, we study the case of a Gaussian random potential
landscape for which we find the universal result of $(3-\sqrt{3})/6 \approx
21\%$ stable area; this type of potential is typically used in the context of
analytical work on random manifolds \cite{HalpinHealy1995} and
disorder-induced pinning \cite{Blatter1994, Giamarchi1995, Nattermann2000}.

The Hessian matrix of random landscapes has been studied in different
contexts, ranging from more abstract discussions of the statistics of critical
points (where gradients vanish) of Gaussian fields in high-dimensional spaces
 \cite{Bray2007, Fyodorov2018} or topological rules for their arrangement in a
random phase field \cite{Freund1995}, to more specific analyses of the
intensity of laser speckle patterns \cite{Weinrib1982} or the complexity of
the free energy function in a model glass \cite{Annibale2003}, see Ref.~\cite{Fyodorov2018} for an extended list of references. Here, we focus
on a planar energy landscape (the pinscape) where we are interested in its
stable area, i.e., the collection of all points that can become minima under
appropriate tilt, rather than studying the (spectral) distribution of
individual critical points (minima, maxima, and saddles).

\begin{figure}[t]
\centering
\includegraphics[width = .45\textwidth]{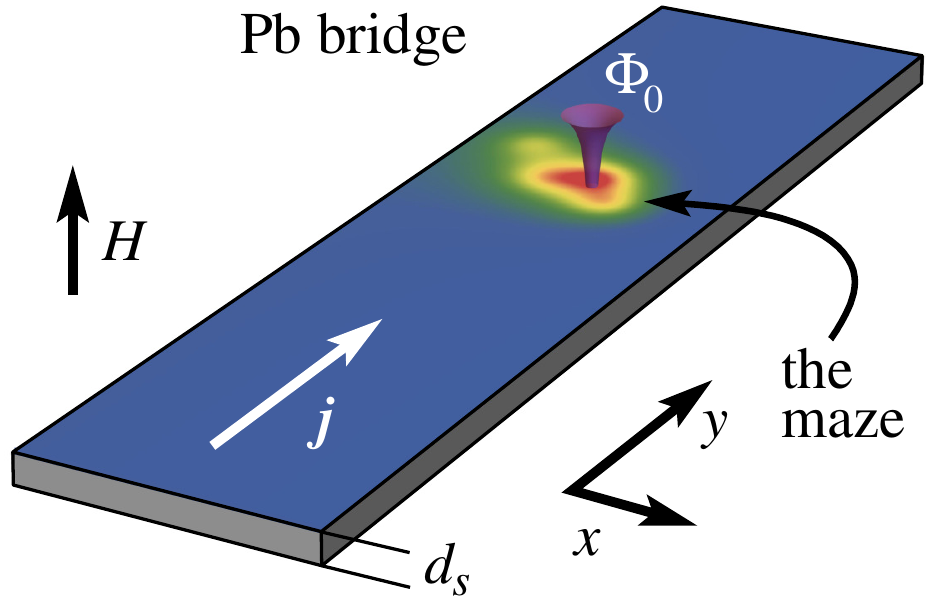}
\caption{Setup for carrying out the `vortex-in-the-maze' experiment inspired
from Ref.~\cite{Embon2015}. A Pb-film of thicknes $d_s$ of order of
the coherence length (and of the same order as the penetration depth) is
subject to an external field $H$ producing vortices in the film. The current
density $j \| y$ drives the vortex along the $x$ direction. Different entry
points along the $y$ direction allow to probe other parts of the pinscape.}
\label{fig:exp_layout}
\end{figure}

In the experiment of Ref.~\cite{Embon2015}, a vortex (carrying a
quantum $\Phi_{0} = hc/2e$ of magnetic flux) is driven across a
two-dimensional superconducting strip made from lead (Pb).  The variations of
the vortex energy across the strip defines the pinning landscape where
the vortex can be trapped in local minima, see Fig.~\ref{fig:exp_layout}---we
refer to this pinscape as `the maze'. These local minima can be manipulated by
applying a transport current $j$ along the constriction (the $y$ direction)
that tilts the potential landscape to the right (in the $x$ direction). In the experiment, a small
$ac$ current imposed on top of the $dc$ drive allows for the precise tracking
of the vortex position.

When drawing a comparison between the dexterity game and the vortex
experiment, few similarities and differences are to be noticed: In both
setups, the ball or vortex can only be stabilized in subregions of the maze,
where, upon applying the proper tilt, the ball or vortex can be trapped in a
local minimum. It is this local minimum which then is manipulated by the
external force, gravity through geometric tilt in the case of the ball, a
transverse current producing the Lorentz force in the case of the vortex.
Tilting the ball's potential beyond the critical slope, the ball rolls along a
guiding plane to the next barrier where its motion stops.  Similarly, pushing
a vortex beyond a region of stable points (as defined by the Hessian of the
potential surface, see below), the vortex crosses the landscape until it gets
retrapped in a suitable local minimum within another stable region.  The two
objects, ball and vortex,  move quite differently, though, with a massive
dynamics $m \ddot{\vec{r}}$ governing the ball's motion, while the vortex
motion is dissipative, $\eta \dot{\vec{r}}$ with $\eta$ denoting the vortex
viscosity
 \cite{Bardeen1965}.

Now, the question may be asked, what regions of the pinscape can be probed at
all, i.e., which points in the plane allow for a local minimum in the (tilted)
potential landscape (or the maze)---this question will take us to the Hessian
stability map of the disorder potential, see Fig.~\ref{fig:stability-region}
below. A quantitative question then is about the total area fraction where a
vortex can be stabilized in a fixed position of the pinning landscape, given
an appropriate (linear) force---this question is addressed by the calculation
of the Hessian character.  While the ball can be driven along both planar axes
$x$ and $y$, subjecting the vortex to a current along $y$, the ensuing Lorentz
force will drive the vortex exclusively along $x$ [see the missing second
control knob in Fig.~\ref{fig:mazes}(b)] with the trajectory running in 2D
plane.  The one-dimensional nature of the trajectory, however, is complicating
the task of mapping out the two-dimensional potential landscape.  One possible
way out is to make use of different `entry points' for the vortex along the
$y$ axis (see Fig.~\ref{fig:exp_layout}) and repeat the `vortex-in-the-maze'
experiment several times---this has been partly (but not systematically) done
in Ref.~\cite{Embon2015}. Another possibility, briefly discussed in this
paper, is to induce a local motion along $y$ with the help of an additional
high-frequency $ac$ drive and measuring the out-of-phase response signal; this
technique allows to expand the probing region in the $y$ direction but may be
quite demanding, depending on the material and experimental parameters.

The reconstruction of vortex tracks in Ref.~\cite{Embon2015} has
brought forward interesting observations in the vortex dynamics at the center
and edge of the Hessian stable regions: for one, a very large $ac$ amplitude
in the middle of the potential well suggests a strong softening of the
confining potential, while the abrupt departure of the vortex from the
defect---with no significant softening and absence of a maximum in the pinning
force---has inspired the 'broken-spring' effect. We will briefly comment on
these features below.

Before entering the discussion of the Hessian map and character, we briefly
discuss in Sec.~\ref{sec:in-the-pin} the pinscape spectroscopy used in the
reconstruction of the pinning landscape \cite{Embon2015}. The definition of the
Hessian stability map in Sec.~\ref{sec:Hessian} then follows quite naturally
and we discuss its various relations to the pinscape spectroscopy of Ref.~\cite{Embon2015}.  In section \ref{sec:char-pinscapes}, we focus on
the main topic of this paper, the determination of the Hessian character of
various types of pinscapes. Section \ref{sec:summary} provides a short
summary.

\section{Pinscape spectroscopy}\label{sec:in-the-pin}

As a motivation to study the Hessian stability map and the Hessian character
of a pinscape, we start with briefly reminding the setup and technique of
Ref.~\cite{Embon2015}, see also Fig.~\ref{fig:exp_layout}, that allows for
mapping out the pinning landscape of vortices in a type-II superconducting
film. Applying a current $\vec{j}\parallel \hat{\vec{y}}$ along the $y$
direction of the film, the total force $\vec{F}$ acting on the vortex involves
the two contributions $\vec{F} = \vec{F}_{\!\mathrm{pin}} +
\vec{F}_{\!\mathrm{\scriptscriptstyle L}}$, where $\vec{F}_{\!\mathrm{pin}} =
- \nabla U(\vec{r})$ accounts for the potential landscape $U(\vec{r})$
[$\vec{r} = (x,y)$ is the two-dimensional coordinate] and
$\vec{F}_{\mathrm{\scriptscriptstyle L}} = \Phi_{0} j d_{s}\hat{\vec{x}}/c$ is
the current-induced Lorentz force, with $d_s$ the film thickness. The Lorentz
force effectively tilts the pinscape $U(\vec{r}) \to
U_{\mathrm{tilt}}(\vec{r}, F_{\mathrm{\scriptscriptstyle L}}) = U(\vec{r}) -
F_{\mathrm{\scriptscriptstyle L}} x$ in the $x$ direction.

Besides the excellent resolution of the SOT device, the precise determination
of the vortex position in the vortex-in-the-maze experiment \cite{Embon2015}
relies on a shaking technique where an additional small oscillatory $ac$
current $j_{ac}\, \exp(-i\omega t)$ is applied on top of the $dc$ drive.
The vortex trajectory $\vec{u}(t)$ then is governed by the dissipative
equation of motion
\begin{align}\label{eq:eom}
   \eta \dot{\vec{u}} = \vec{F}(\vec{u},t),
\end{align}
with $\eta$ the viscosity and $\vec{F} = \vec{F}_{\!\mathrm{pin}} +
\vec{F}_{\mathrm{\scriptscriptstyle L}} + \vec{F}_{ac}\, \exp(-i\omega t)$ the
total force acting on the vortex. By applying a sequence of increasing $dc$
tilts $F_{\mathrm{\scriptscriptstyle L}n}$, the vortex will move forward
through the pinscape and oscillate around a tilt-dependent minimum
$\vec{r}_{n}(F_{\mathrm{\scriptscriptstyle L}n})$. Near this position, the
associated energy profile can be expanded in the displacement $\vec{u} =
\vec{r} - \vec{r}_{n}$,
\begin{align}\label{eq:tilt-potential}
   \!\!\!
   U_{\mathrm{tilt}}(\vec{r},F_{\mathrm{\scriptscriptstyle L}n})
   &\!=\! U_{\mathrm{tilt}}(\vec{r}_{n},F_{\mathrm{\scriptscriptstyle L}n}) 
   + a_{n} u_{x}^{2}\!+ b_{n} u_{y}^{2}\! + c_{n} u_{x} u_{y},\!
\end{align}
with higher-order corrections becoming relevant near the edges of the stable
regions. The local curvatures $a_{n} = a(\vec{r}_{n})$, $b_{n} =
b(\vec{r}_{n})$, and $c_{n}= c(\vec{r}_{n})$ define the Hessian matrix via
Eq.~\eqref{eq:Hessian-matrix}, see below.

Expressing the vortex displacement through $\vec{u} = (u_{x},u_{y})\,
e^{-i\omega t}$, the equation of motion \eqref{eq:eom} takes the form
\begin{align}\label{eq:ux0}
   i \eta \omega u_{x} &= 2a u_{x} + c u_{y} - F_{ac},\\
   i \eta \omega u_{y} &= 2b u_{y} + c u_{x}.\label{eq:uy0}
\end{align}
These equations can be solved and analyzed perturbatively in the small
parameter $\eta\omega/U''$ involving the viscous term $\eta\omega$ and the
curvatures $U'' \sim a,\, b,\, c\,$; indeed, simple estimates (see Appendix
\ref{app:parameter}) show that this ratio is small for the material and
setup in Ref.~\cite{Embon2015}.  Solving Eqs.~\eqref{eq:ux0} and
\eqref{eq:uy0} and expanding the result to lowest (0-th) order in
$\eta\omega/U''$, we find that
\begin{align}\label{eq:zeroth-order-solution}
   u_{x} = \frac{F_{ac}}{2 a (1-c^2/4ab)}
   \quad \text{and} \quad
   u_{y} = (-c / 2b) u_{x}.
\end{align}
The motion is in phase with the external
driving force and follows the local potential minimum. Hence, although the
$ac$ force is applied along $x$, the vortex oscillates at a finite angle $\phi
= \arctan(u_{y}/u_{x}) = -\arctan(c/2b)$ away from the $x$ axis in the
direction of its trajectory.

Given the displacement amplitudes $u_x$ and $u_y$, one easily reconstructs the
potential along the vortex trajectory.  For the specific choice of linear
increments $F_{\mathrm{\scriptscriptstyle L}n} = n F_{ac}$ \cite{Embon2015},
the equilibrium position $\vec{r}_{n}$ at the drive
$F_{\mathrm{\scriptscriptstyle L}n}$ relates to the position $\vec{r}_{n-1}$
via
\begin{align}\label{eq:iterative-displacement}
   \vec{r}_{n} = \vec{r}_{n-1} + (u_{x, n-1}, u_{y, n-1}),
\end{align}
where $u_{x, n}$, $u_{y, n}$ are the $ac$ displacement amplitudes
\eqref{eq:zeroth-order-solution} measured at the drive
$F_{\mathrm{\scriptscriptstyle L}n}$. This trivial iterative relation leads to
the trajectory $\vec{r}_{n} = \sum_{m=0}^{n-1}(u_{x, m}, u_{y, m})$.
Combining the definition of the tilted potential $U_{\mathrm{tilt}}$ at
$F_{\mathrm{\scriptscriptstyle L}n} = n F_{ac}$ with the quadratic
approximation \eqref{eq:tilt-potential}, we obtain
\begin{align}\label{eq:tilt-again}
   U_{\mathrm{tilt}}^{n}(x,y) &= U(x,y) - n F_{ac} x\\
   \nonumber
      &\approx U_{\mathrm{tilt}}^{n}(x_{n},y_{n}) + a_{n} (x - x_{n})^{2}
      + b_{n} (y - y_{n})^{2}\\
      \nonumber
      &\qquad\qquad\qquad\;\;  + c_{n} (x - x_{n})(y - y_{n}).
\end{align}
Solving for $U(x_{n-1},y_{n-1})$ and $U(x_{n},y_{n})$ and combining the
results with Eqs.~\eqref{eq:zeroth-order-solution} and
\eqref{eq:iterative-displacement}, one finds the change in the pinning
potential between neighboring points (we choose the arbitrary offset
$U(\vec{r}_{0}) = 0$),
\begin{align}\label{eq:iterative-potential}
   U(x_{n},y_{n})
     &\approx U(x_{n-1},y_{n-1}) + (n - 1/2)\,u_{x, n-1} F_{ac}
\end{align}
and its iteration provides us with potential
\begin{align}\label{eq:equil-potential}
   U(\vec{r}_{n}) \approx F_{ac} \sum\limits_{m=0}^{n-1} (m + 1/2)\, u_{x,m}.
\end{align}
The reconstruction of the pinscape along a trajectory in 2D only involves a 1D
integral along $x$, a consequence of the unidirectional tilt.  Indeed, the
implicit stability criterion along $y$, $\partial U / \partial y = 0$, reduces
the integration in the $xy$ plane to the simple 1D form of Eq.~\eqref{eq:iterative-potential}.

The above scheme allows for the reconstruction of the pinscape along the
trajectory. Interestingly, the solution and subsequent expansion of Eqs.~\eqref{eq:ux0} and \eqref{eq:uy0} to linear order in $\eta \omega/U''$
provides an out-of-phase correction $\delta u_x, \, \delta u_y \propto
i(\eta\omega /U'')/U''$ that could be measured independently, at least in
principle.  The four displacements $u_x$, $u_y$, $\delta u_x$, and $\delta
u_y$ then allow for the determination of all local curvatures $a$, $b$, and
$c$ and thus give access to the local reconstruction of the potential $U(x,y)$
within a strip around the trajectory; details of this extension of pinscape
spectroscopy are presented in Appendix \ref{app:1st-order}.

\section{Hessian Stability Map}\label{sec:Hessian}

Given the possibility to map out the pinning potential of a film through
pinscape spectroscopy, the question poses itself which part of the plane can
actually be analyzed in this manner and what happens at the boundaries of
these areas; the answer to these questions is given by the Hessian stability
map.

In the absence of an $ac$ current, the vortex resides in a minimum of the
tilted potential $U_{\mathrm{tilt}}(\vec{r}, F_{\mathrm{\scriptscriptstyle
L}})$. Such a \emph{stable point} is characterized by a vanishing first
derivative along both $x$ and $y$ (no net force) and a positive curvature. The
second condition is satisfied, if the Hessian matrix
\begin{align}\label{eq:Hessian-matrix}
   H(x,y) &= \begin{pmatrix}
   \frac{\partial^{2} U}{\partial x^{2}} &
   \frac{\partial^{2} U}{\partial x \partial y}\\
   \frac{\partial^{2} U}{\partial y \partial x} &
   \frac{\partial^{2} U}{\partial y^{2}}
   \end{pmatrix}
   = \begin{pmatrix}
   2a(x,y) &
   c(x,y)  \\
   c(x,y)  &
   2b(x,y)
   \end{pmatrix}
\end{align}
is positive-definite, i.e., it has a positive determinant 
\begin{align}\label{eq:stability-Hesse}
   \det H(x,y) =  4a(x,y)b(x,y) - c^2(x,y) > 0
\end{align}
and a positive trace 
\begin{align}\label{eq:stability-Hesse-tr}
   \tr H(x,y) =  2[a(x,y) + b(x,y)] > 0.
\end{align}
Here, the coefficients $a(\vec r)$, $b(\vec r)$, and $c(\vec r)$ coincide with
the local expansion coefficients in Eq.~\eqref{eq:tilt-potential}. While
Eq.~\eqref{eq:stability-Hesse} only excludes indefinite matrices (saddle-point
solutions), the positive trace \eqref{eq:stability-Hesse-tr} discards
negative-definite Hessian matrices (potential maxima).  Note that the Hessian
does not depend on the (linear) drive, hence it characterizes the pinscape
$U(x,y)$ itself, rather than the forced pinscape $U_\mathrm{tilt}$.  As such,
the Hessian matrix with its determinant and trace provides information on the
potential's capability of stabilizing a vortex at a specific point $\vec{r}$
of the plane upon application of the appropriate tilt.
\begin{figure}[tb]
\centering
\includegraphics[width = 7truecm]{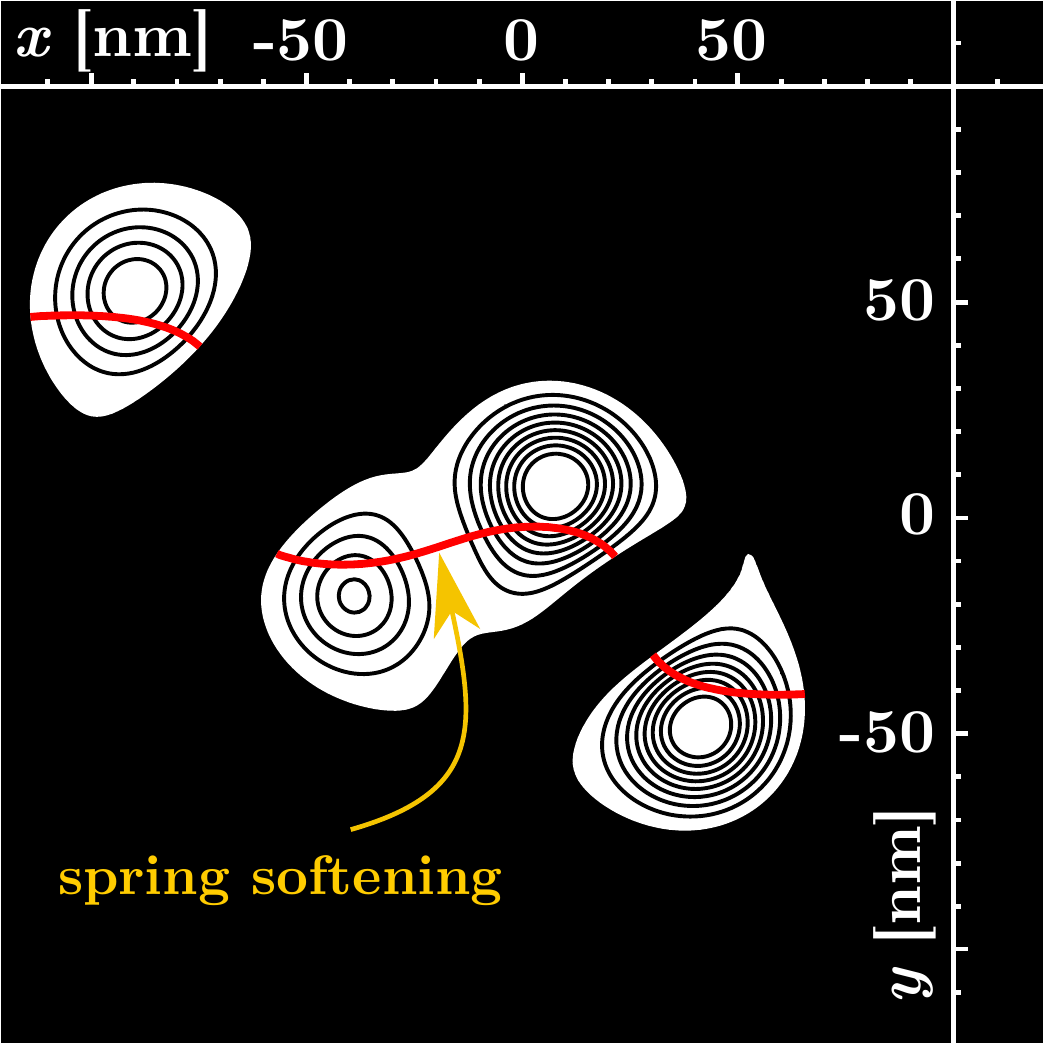}
\caption{A new view on the pinning landscape through the Hessian stability
map.  Shown is the example of the pinscape derived from measurements in
Ref.~\cite{Embon2015}, see Fig.~5(a) therein, and shown on the game-board in
Fig.~\ref{fig:mazes}(b).  All of the black area is unstable, i.e., the Hessian
matrix Eq.~\eqref{eq:Hessian-matrix} has at least one negative eigenvalue;
vortices cannot be trapped at any point within this region and the potential
landscape cannot be probed. A position within the white area is stable and
turns into a local minimum for a specific tilt along $x$ and $y$.  Contour
lines show equi-Hessians where $\det H = 0.3 k\, (\mathrm{meV/nm^{2}})^{2}$
for integer $k$. For a unidirectional tilt along $x$, only one specific
trajectory (red) is accessible within the stable regions.  Close to the border
of the stable regions, the Hessian becomes small and the $ac$ response of the
vortex increases. The divergence of the $ac$ displacement at the Hessian
boundary is preempted by the thermal activation out of the well and subsequent
run-away of the vortex across the unstable region. At the center of the
double-defect (yellow arrow) the vortex goes though a flat region with a small
Hessian, implying a large $ac$ response (spring softening) as observed in
Fig.~2(e) of Ref.~\cite{Embon2015}.} \label{fig:stability-region}
\end{figure}

The traditional way of studying the potential landscape is via equipotential
(or elevation) maps. They depend on the current-induced tilt and their minima
tell about possible (meta-)stable positions for the vortex. Adopting a global
view, the Hessian matrix helps separating stable points from unstable points.
This way, the two-dimensional pinning landscape can now be divided into stable
areas characterized by the set of conditions $\det H(x,y) > 0$ and $\tr H(x,y)
> 0$, and unstable ones where at least one condition is violated. We thus
introduce the \emph{Hessian stability map}, i.e., the graphical representation
of the pinscape regions associated with stable points, as a new tool to
characterize a potential landscape, with a `good' pinscape described by a
large percentage of stable area.  In Fig.~\ref{fig:stability-region}, we show,
for illustration, the stability region, together with equi-Hessian contour
lines, for the potential landscape considered in Ref.~\cite{Embon2015} [see
Fig.~5(a) therein] and also shown on the game-board in Fig.~\ref{fig:mazes}(b)
as well as the setup in Fig.\ \ref{fig:exp_layout}. Within the black regions,
at least one eigenvalue is negative, implying that this position cannot be
made a stable vortex position for any tilt (in either $x$ and $y$ direction).
In the following, we briefly discuss the role played by the Hessian map in the
context of pinscape spectroscopy via $ac$ and $dc$ forces. In Section
\ref{sec:char-pinscapes}, we assume a more generic view on the problem and
determine the Hessian character, i.e., the area fraction of stable regions,
for different potential landscapes often used in numerical or analytical
studies of vortex pinning and dynamics.  These are a finite density of cut
parabolas, of Gaussian and Lorentzian shaped potentials, as well as a Gaussian
random potential.

Let us first interpret the Hessian stability map and extract some physical
insights into the pinscape.  Focusing on the boundaries in the stability map,
we note that the vortex displacement $\vec{u} \propto (1-c^2/4ab)^{-1}$
diverges, as $c^{2} \!\to\! 4ab$ when the minimum in $\vec{r}_{n}$ approaches
the boundary, see Eq.~\eqref{eq:zeroth-order-solution}.  Upon approaching the
singular point $c^{2} = 4ab$, the expression for the trajectory's angle $\phi$
simplifies to $\phi = \arctan[(a/b)^{1/2}]$ and thus provides access to the
ratio of the potential curvatures along the directions $x$ and $y$.
Interesting features show up when multiple defects combine into a more complex
pinning landscape \cite{Embon2015}. For example the vortex can approach the
depinning point of one defect and transit to another without entering the
unstable region of the pinscape. The pinscape then develops a flat region with
a small Hessian determinant in the middle of the well. As a result, the $ac$
displacement amplitude rises steeply as observed in Ref.~\cite{Embon2015},
what corresponds to a \emph{spring softening} as highlighted in
Fig.~\ref{fig:stability-region}, yellow arrow. Analyzing the vortex trajectory
in the central defect more carefully, one notes that the vortex traverses
(from left to right) a region going from $\det H \!\sim\!  1$ (meV/nm$^2)^2$
near the first minimum, to a small value below $0.3$ (meV/nm$^2)^2$ near the
`saddle', to a large value $\sim\! 2$ (meV/nm$^2)^2$ in the second minimum.
One thus expects an enhancement of the $ac$ amplitude by a factor of 3--4
starting at the left of the spring softening and a factor 7--8 relative to the
value at the right side, in qualitative agreement with the experiment.

Another aspect of interest is the escape of the vortex from the stable regime.
The proper understanding of this phenomenon requires to include higher-order
terms in the local expansion of the potential $U(x,y)$ and involves thermal
escape over barriers and possibly anharmonic effects, see Appendix
\ref{app:escape} for details. Our semi-quantitative analysis of the setup in
Ref.~\cite{Embon2015} confirms that thermal fluctuations are strong and
trigger the escape of the vortex from the stable region at quite a large
distance away from the stability boundary, in agreement with the discussion of
the `broken-spring effect' in the experiment. Specifically, thermal
fluctuations and vortex escape do cut off the expected divergence in the
displacement $\vec{u} \propto (1-c^2/4ab)^{-1}$ and the reconstructed pinning force
does not go through a maximum at the point of escape.

\section{Hessian character of pinscapes}\label{sec:char-pinscapes}

We now turn to the main part of this paper, the calculation of the Hessian
character $\mathcal{C}_{\mathrm{pos}}$ of a pinscape. This number quantifies
the fraction (less than unity) of the plane's area that admits a stable vortex
position (i.e., a positive definite Hessian matrix) under an appropriate tilt
force. The Hessian has been used in the characterization of various functions
in a multitude of fields, including optics \cite{Weinrib1982,Freund1995},
statistical physics of random systems
 \cite{Annibale2003,Bray2007,Fyodorov2018}, or cosmology \cite{Yamada2018}, see
Ref.~[\onlinecite{Fyodorov2018}] for a more detailed list of references.
Those studies typically focus on a set of specific critical points in a given
area (corresponding to extremal points at a given fixed tilt in the present
context), while we aim at characterizing every point in space as potentially
giving rise to a minimum under an appropriate tilt. As a result, here, we
determine the area fractions with specific curvature properties.

A point $\vec{r} \!\in\! \Omega$ in the two-dimensional landscape of area $\Omega$
is called stable if the local potential landscape features a positive-definite
Hessian matrix; the collection of such stable positions defines the stability
regions of the pinscape where the pinscape can be mapped through the
spectroscopic method described in Sec.~\ref{sec:in-the-pin}. 

\subsection{Single defect}\label{sec:single-defect}

As a warmup, consider the pinscape of a single defect. Here, we focus on
isotropic defects with a potential $V(\vec{r}) = V(r)$, assuming a minimum
$-V_0$ at the origin $r = 0$, and a monotonic radial dependence $V'(r) > 0$,
where the prime $'$ denotes the radial derivative $V'(r) = \partial_r
V(r)$.  We demand the potential to be integrable, $\int d^{2}{r}\,
|V(\vec{r})| < \infty$, implying its asymptotic decay $V(r \!\to\!  \infty) =
0$; a notable exception is the long-range Lorentzian potential discussed
below. The Hessian matrix of such an isolated defect possesses the eigenvalues
$V''(r)$ and $V'(r) / r$; they describe longitudinal (along $\vec{r}$) and
transverse (to $\vec{r}$) curvatures. While the latter is positive everywhere,
the longitudinal curvature assumes a positive value only in the vicinity
of the defect's center. Defining the stability radius $\xi_{0}$ through the
condition $V''(r = \xi_{0}) = 0$, we find the stable area $\Omega_{0} = \pi
\xi_{0}^{2}$; at distances larger than $\xi_0$, the landscape is indefinite.
Maxima appear in the pinscape only through the interference of (at least two)
defects.

For the specific cases of a Gaussian-shaped
\begin{equation}\label{eq:V_G}
V_{\mathrm{\scriptscriptstyle G}}(r) =  -V_{0} \exp(-r^{2}/\xi^{2}) 
+ \bar{V}_{\mathrm{\scriptscriptstyle G}}
\end{equation}
and Lorentzian-shaped
\begin{equation}\label{eq:V_L}
V_{\mathrm{\scriptscriptstyle L}}(r) = -V_{0}/(1+r^{2}/\xi^{2}) 
+ \bar{V}_{\mathrm{\scriptscriptstyle L}} 
\end{equation}
defect potential, we find the stability radii $\xi_{0} \!=\! \xi/\sqrt{2}$ and
$\xi_{0} \!=\! \xi/\sqrt{3}$, respectively.  The constant shifts
$\bar{V}_{\mathrm{\scriptscriptstyle G}} \!=\! V_0 \pi\xi^{2}/\Omega$ and
$\bar{V}_{\mathrm{\scriptscriptstyle L}} \!=\!  V_0 (\pi\xi^{2}/\Omega) \ln[1
+ (\Omega/\pi\xi^{2})]$ assure a vanishing potential average, i.e.,
$\int_{\Omega} d^{2}r\, V(\vec{r}) \!=\! 0$. Below, we will also consider the
case of a cut parabola 
\begin{equation}\label{eq:V_P}
V_{\mathrm{\scriptscriptstyle P}}(r) =
-V_{0}(1-r^2/\xi^2) \Theta(r-\xi) + \bar{V}_{\mathrm{\scriptscriptstyle P}},
\end{equation}
with $\xi_0 \!=\!  \xi$ and $\bar{V}_{\mathrm{\scriptscriptstyle P}} \!=\!
V_{0} \pi \xi^{2}/2\Omega$; this type of potential has often been used in
numerical simulations of vortex pinning \cite{Reichhardt1995, Olson2017}.

Next, we consider a pinscape originating from a small density $n_p = N/\Omega$
of defects, where $N$ denotes the number of defects in the area $\Omega$. For
a very low density of defects, $n_{p} \Omega_{0} \ll 1$, the probability $\sim
(n_{p}\xi^{2})^{2}$ for defects to overlap is parametrically small; as a
result the stability region to leading order in $n_{p}\xi^2$ assumes the value
\begin{align}\label{eq:C_dilute}
   \mathcal{C}_{\mathrm{pos}} \approx n_{p} \Omega_{0}.
\end{align}
This generic result tells, that only a minute areal fraction in the immediate
vicinity of defects is capable of being probed within the vortex-in-the-maze
scheme.

\subsection{Gaussian limit of dense defects}\label{sec:Gaussian}

The nontrivial and hence interesting structure of a pinscape develops when
defect potentials start to overlap. Below, we study pinning landscapes of the type
\begin{align}\label{eq:Ur}
   U(\vec{r}) = \sum\nolimits_{j=1}^{N} V(\vec{r}-\vec{r}_{i}).
\end{align}
We assume $\int_{\Omega} d^{2}r \, V(\vec{r}) = 0$ such that the potential $U$
averages to zero as well. Given a random distribution of defect positions
$\vec{r}_i$, the pinscape turns into a random energy surface. Our task now
consists in determining the (mean) character $\mathcal{C}_{\mathrm{pos}}$ for
specific types of random landscapes. The latter is defined through the
probability density $p(\D,\T)$ of finding a position with given Hessian
determinant $\det H \!=\! \D$ and trace $\tr H \!=\! \T$, both of which have
to be positive $\D > 0$ and $\T > 0$,
\begin{align}\label{eq:C}
   \mathcal{C}_{\mathrm{pos}} = \int_{0}^{\infty} \!\!\!\int_{0}^{\infty}\!
   d\D \, d\T\ p(\D,\T).
\end{align}
Characterizing the random pinscape potential $U(\vec{r})$ through its
functional probability measure $\mathcal{P}[U(\vec{r})]$, we find the
probability density $p(\D,\T)$ via functional integration,
\begin{align}\label{eq:pXY}
   p(\D,\T) = \! \int \! \mathcal{D}[U(\vec{r})]\, \mathcal{P}[U(\vec{r})]\,
   \delta[\det H - \D] \, \delta[\tr H - \T],
\end{align}
where the Hessian matrix $H$ can be evaluated at any spatial point
$\vec{r}$ due to the translation invariance of the result; without loss of
generality, we choose $\vec{r} = \vec{0}$. For a homogeneous distribution of
$N$ defects in an area $\Omega$, see Eq.~\eqref{eq:Ur}, the measure in
\eqref{eq:pXY} is given by
\begin{align}\label{eq:DUr}
   \mathcal{D}[U(\vec{r})]\, \mathcal{P}[U(\vec{r})]
   = \prod_{j=1}^{N} \Big[\frac{d^{2} r_{j}}{\Omega}\Big].
\end{align}

A second generic result [besides the trivial dilute limit \eqref{eq:C_dilute}]
can then be obtained in the high density limit $n_p\Omega_0 \gg 1$ when many
defects overlap.  As shown in Appendix \ref{app:central-limit}, the pinscape
of many overlapping defects approaches a Gaussian distribution with vanishing
mean $\langle U(\vec{r}) \rangle = 0$ [since $\langle V(\vec{r}) \rangle = 0$]
and a two-point correlator
\begin{align}\label{eq:correlator}
   \!\!
   G(\vec{r} \!-\! \vec{r}') \!=\! \langle U(\vec{r}) U(\vec{r}') \rangle
   \! = \! n_{p} \! \!\int\!\! d^{2} s \, V(\vec{r} \!-\! \vec{s}) 
                                           V(\vec{r}' \!-\! \vec{s})
\end{align}
deriving from the convolution of two shifted potentials $V(\vec{r})$. It
follows from the central limit theorem that the distribution function
$\mathcal{P}[U(0)]$ for the potential in a fixed point, e.g., at the origin,
is of Gaussian form. The fact that the {\it functional} distribution function
$\mathcal{P}[U(\vec{r})]$ becomes Gaussian as well,
\begin{align}\label{eq:Gaussian-measure}
   \mathcal{P}[U(\vec{r})] = \mathcal{P}_{\mathrm{\scriptscriptstyle G}}[U(\vec{r})] =
   e^{-\mathcal{S}} / \mathcal{Z},
\end{align}
with $\mathcal{Z} = \int \mathcal{D}[U(\vec{r})]\ e^{-\mathcal{S}}$ and the
quadratic action
\begin{align}\label{eq:Gaussian-action}
   \mathcal{S} = \frac{1}{2} \int\! \frac{d^{2} r}{\Omega}\! \int\!
   \frac{d^{2} r'}{\Omega}\ U(\vec{r}) \, G^{-1}(\vec{r} - \vec{r}')\, U(\vec{r}'),
\end{align}
is less trivial and can be checked by confirming the validity of Wick's
theorem for the $2k$-point correlators (up to corrections in the small
parameter $1/n_p\Omega_0$) or via a direct calculation of
$\mathcal{P}[U(\vec{r})]$, see Appendix \ref{app:central-limit}.

For such a Gaussian random potential, symmetry imposes that regions of
positive- and negative-definite Hessians (i.e., with $\D \!>\! 0$ and
$\sign(\T) \!=\! \pm 1$ respectively) are equally
probable and hence Eq.~\eqref{eq:C} reduces to the evaluation of the
simpler expression
\begin{align}\label{eq:stable-fraction}
   \mathcal{C}_{\mathrm{pos}} = \frac{1}{2}\int_{0}^{\infty} \! d\D\ p(\D),
\end{align}
where $p(\D)$ denotes the probability distribution of the Hessian determinant
$\det H$ taking the value $\D$. 

The task of finding the probability density $p(\D)$ can be broken up into a
sequence of problems: in a first step, we can determine the probability
$\pi(a,b,c)$ for a Hessian matrix to assume diagonal entries $2a$, $2b$ and
off-diagonal entries $c$, thereby reducing the problem of evaluating
Eq.~\eqref{eq:stable-fraction} to an algebraic integral,
\begin{align}
    \label{eq:X-prob-ABC}
    p(\D) &=\!\! \int \! da \, db \, dc \,\pi(a,b,c)\, \delta[4ab-c^{2} - \D].
\end{align}
We find the probability function $\pi(a,b,c)$ via the functional integration
\begin{align}\label{eq:pi-ABC}
\!\!
   \pi(a,b,c) &\!= \!\!\int \!\! \mathcal{D}[U(\vec{r})] \, 
   \mathcal{P}_\mathrm{\scriptscriptstyle G} [U(\vec{r})]
   \, 4\, \delta[U_{xx}(0) - 2a] \\[-.3em]\nonumber
   &\hspace{6em} \times \delta[U_{yy}(0) - 2b]\, \delta[U_{xy}(0) - c].
\end{align}
The numerical factor $4$ appears from applying the identity
$\delta[U_{xx}(0)/2 - a] = 2\, \delta[U_{xx}(0) - 2a]$ and equally for
$\delta[U_{yy}(0)/2 - b]$. The difficulty with the functional integration
over all realizations $U(\vec{r})$ is now moved to the evaluation of
$\pi(a,b,c)$ in Eq.~\eqref{eq:pi-ABC}.

Substituting Eq.~\eqref{eq:Gaussian-action} into Eq.~\eqref{eq:pi-ABC} and
expressing the $\delta$ distributions in Fourier space, we have to the evaluate
\begin{align} \label{eq:pi_Gauss}
   \pi(a,b,c) &= \! \frac{1}{\mathcal{Z}} \! \int \! \mathcal{D}[U(\vec{r})]
   e^{-\frac{1}{2} \int\! \frac{d^{2} r}{\Omega}\! \int\! \frac{d^{2} r'}{\Omega}
      U(\vec{r}) G^{-1}(\vec{r} - \vec{r}') U(\vec{r}')}
   \\[-1em]
   \nonumber
    & \qquad
   \times \int \frac{dk\, dl\, dm}{(2\pi)^{3}} 4 e^{i(2ka + 2lb + mc)}
   \\ \nonumber
    &\quad\qquad
     \times e^{-i \int\! d^{2} r [k U_{xx}(\vec{r}) + l U_{yy}(\vec{r})
      + m U_{xy}(\vec{r})]\delta(\vec{r}) }.
\end{align}
Two integrations by parts in the exponent of the last factor yield $\int\! 
d^{2} r \, U(\vec{r})[k \delta_{xx}(\vec{r}) \!+\! l \delta_{yy}(\vec{r})
\!+\! m \delta_{xy}(\vec{r})]$, with $\delta_{\kappa \mu}(\vec{r}) \!\equiv\!
\partial^{2} \delta(\vec{r}) / \partial x_{\kappa} \partial x_{\mu}$.
The remaining functional integration can now be
performed through Gaussian integration \cite{ZinnJustin2005, Altland2010}
(i.e., completing the square),
\begin{align}\label{}
   \pi(a,b,c) &= \int \frac{dk\, dl\, dm}{(2\pi)^{3}} 4 e^{i(2ka + 2lb + mc)} \\
   \nonumber
   &\hspace{2.5em} \times e^{-\frac{1}{2} [k^{2} G_{0}^{xxxx} + l^{2} G_{0}^{yyyy}
   + (m^{2} + 2kl) G_{0}^{xxyy}]},
\end{align}
where $G_{0}^{\kappa\mu\nu\sigma} \!\equiv\! \partial^{4} G(\vec{r})/ \partial
x_\kappa \partial x_\mu \partial x_\nu \partial x_\sigma |_{\vec{r} = 0} \!>\!
0$ denotes the fourth derivative of the Green's function. For an isotropic
problem, symmetry tells that $G_{0}^{\scriptscriptstyle (4)} \equiv
G_{0}^{xxyy} = G_{0}^{xxxx}/3 = G_{0}^{yyyy}/3$ and hence
\begin{align}\label{}
   \pi(a,b,c) &= \int \frac{dk\, dl\, dm}{(2\pi)^{3}} 4 e^{i(2ka + 2lb + mc)} \\
   \nonumber
   &\qquad \times e^{-\frac{1}{2} [3k^{2} G_{0}^{(4)} + 3 l^{2} G_{0}^{(4)}
   + (m^{2} + 2kl) G_{0}^{(4)}]}.
\end{align}
The remaining Gaussian integrations over $k$, $l$, and $m$ then yield the
result
\begin{align}\label{eq:pi-ABC-res}
   \pi(a,b,c) = \frac{4 e^{-(3a^{2} - 2a b + 3b^{2} + 2c^{2})/4 G_{0}^{(4)}}}
   {\big(4\pi G_{0}^{(4)}\big)^{3/2}}
\end{align}
and we find that the probability distribution of Hessian matrix elements is
Gaussian, as one might have expected for a Gaussian distributed random potential.

Making use of the result \eqref{eq:pi-ABC-res} in Eq.~\eqref{eq:X-prob-ABC},
we find the distribution
\begin{align}\label{eq:pX1}
   p(\D) &= \frac{4}{\big[4\pi G_{0}^{(4)}\big]^{3/2}} \! \int \! \frac{dQ}{2\pi}
   \int \! da\, db\, dc  \, e^{i Q \D} \\ \nonumber
   &\hspace{4em} \times
   e^{-iQ (4ab-c^{2})}
           e^{-(3a^{2} - 2a b + 3b^{2} + 2c^{2})/4G_{0}^{(4)}},
\end{align}
which after another series of Gaussian integrations gives
\begin{align}\label{eq:integral-problem}
   p(\D) = \int\limits_{-\infty}^{\infty}\! \frac{d\tilde{Q}} {2\pi
   G_{0}^{\scriptscriptstyle (4)}} \frac{e^{i\tilde{Q}
   \D/G_{0}^{\scriptscriptstyle (4)}}} {\sqrt{\big(1 -
   2i\tilde{Q}\big)^{2}\big(1 + 4i\tilde{Q}\big) }},
\end{align}
with $\tilde{Q} \!=\! Q G_{0}^{\scriptscriptstyle (4)}$.
\begin{figure}[t]
\centering
\includegraphics[width = .35\textwidth]{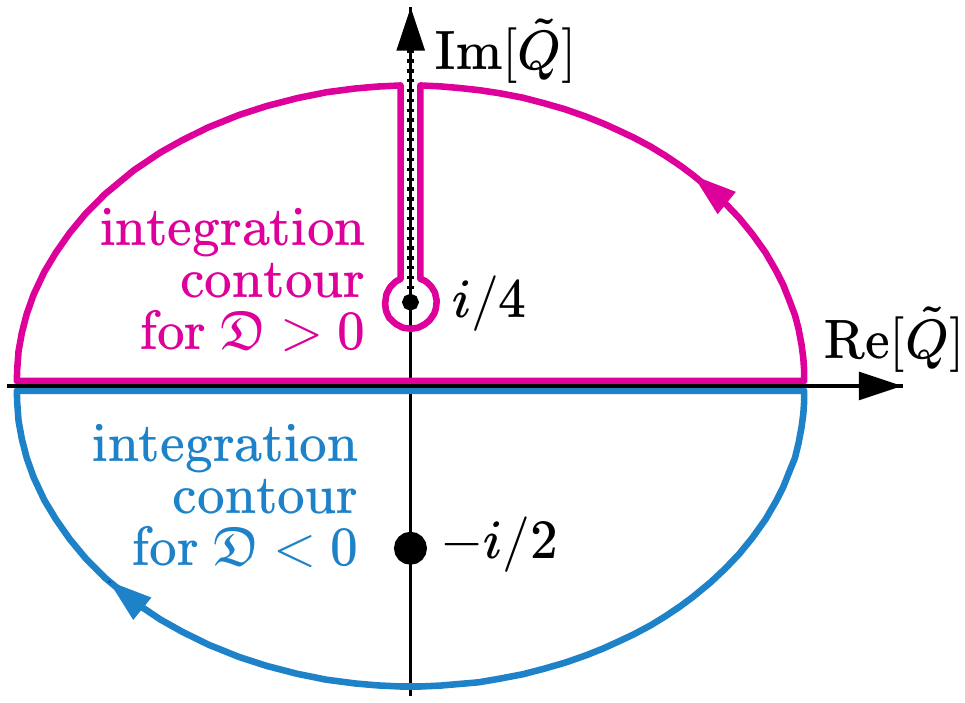}
\caption{Sketches of the contours in the complex plane for evaluating integral
in Eq.~\eqref{eq:integral-problem}. For $\D \!>\! 0$, the (magenta)
contour encloses the upper half-plane except for the cut along the imaginary
axis starting from $i/4$. For $\D \!<\! 0$, the (blue) contour encloses the
lower half-plane with a pole at $-i/2$.  } \label{fig:complex-plane}
\end{figure}
The integrand has a pole of order one in the negative complex plane at
$\tilde{Q} \!=\! -i/2$ and a line cut along the positive imaginary axis,
terminating at $\tilde{Q} \!=\! i/4$, see Fig.~\ref{fig:complex-plane}. The above
integral can be solved for $\D \!>\! 0$ using a closed contour in the upper
complex plane avoiding the line cut along the imaginary axis.
We then find with the substitution $\zeta = \arccot[(4q/3)^{1/2}]$
\begin{align}\label{eq:probability-distribution-Dlarger0}
\!\!\!
   p(\D \!>\!0)
   &= \frac{2 e^{-\D/4G_{0}^{(4)}}}{G_{0}^{(4)}}
   \int_{0}^{\infty} \frac{dq}{2\pi}\frac{e^{-q \D/G_{0}^{(4)}}}{(3+4q)\sqrt{q}}\\
   &= \frac{2 e^{\D/2G_{0}^{(4)}}}{\sqrt{3} G_{0}^{(4)}}
   \int_{0}^{\pi/2} \frac{d\zeta}{2\pi}e^{-\big(3\D / 4G_{0}^{(4)}\big) (\sin\zeta)^{-2}}.\!\!
\end{align}
The integral in the last line is Craig's formula \cite{Craig1991} for the
complementary error function $\Erfc[z] \equiv 1 - \Erf[z]$ for non-negative $z
\!=\! (3\D / 4G_{0}^{(4)})^{1/2}$, with the error function defined as $\Erf(z)
\!=\! (4/\pi)^{1/2} \int_{0}^{z}dt\,e^{-t^{2}}$.
For $\D < 0$ the contour is closed in the lower half-plane, encircling the
pole at $\tilde{Q} = -i/2$. The residue theorem then yields $p(\D \!<\!0)
\!=\! e^{\D/2G_{0}^{\scriptscriptstyle (4)}} /
(2\sqrt{3}G_{0}^{\scriptscriptstyle (4)})$.

\begin{figure}[t]
\centering
\includegraphics[width = .40\textwidth]{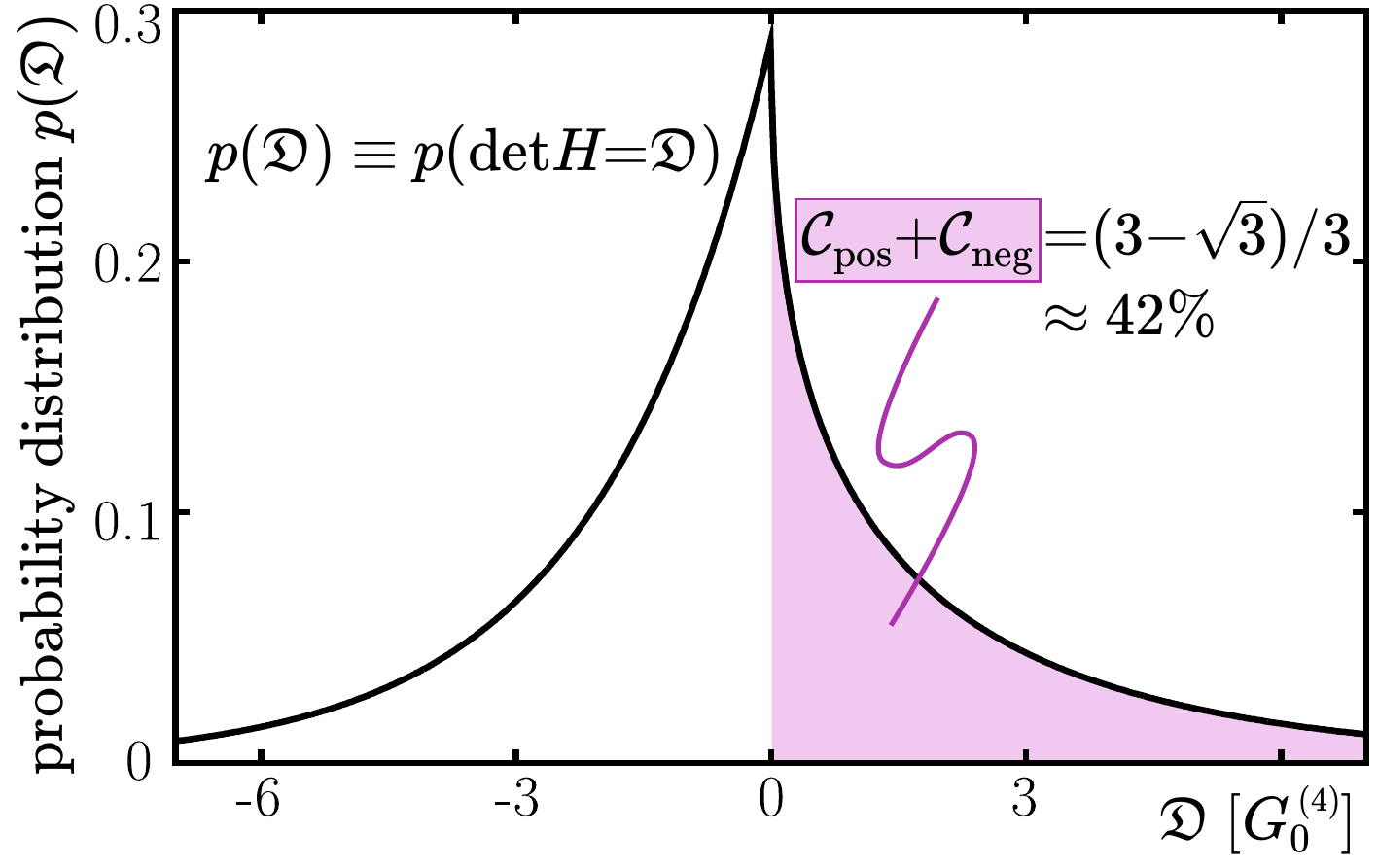}
\caption{Probability distribution function $p(\D)$ of the Hessian $\det H$ for
a Gaussian distributed random potential, see
Eq.~\eqref{eq:probability-distribution-4-Gaussian}. The horizontal axis
measures the determinant $\D$ in units of $G_{0}^{\scriptscriptstyle (4)}$.
The shaded probability indicates the area-fraction of points with positive- or
negative-definite curvature.}
\label{fig:probability-distribution}
\end{figure}

The probability distribution $p(\D)$ for the Hessian determinant then takes
the compact global form (see Fig.~\ref{fig:probability-distribution} for an
illustration)
\begin{align}\label{eq:probability-distribution-4-Gaussian}
\!\!
   p(\D) = \frac{e^{\D/2G_{0}^{\scriptscriptstyle (4)}}}
   {2\sqrt{3}G_{0}^{\scriptscriptstyle (4)}}
   \bigg[1 - \Erf\bigg(\sqrt{\frac{3}{4} 
   \frac{\D}{G_{0}^{\scriptscriptstyle (4)}}}\, \bigg)
   \Theta(\D/G_{0}^{\scriptscriptstyle (4)}) \bigg],
\end{align}
where we have expressed the result through the Heaviside function $\Theta(z)
\!=\! 1$ for $z \!>\! 0$ (and zero otherwise).  The result behaves as $p(\D)
\!\approx\!  p(0)[1-(3\D/\pi G_{0}^{\scriptscriptstyle (4)})^{1/2}]$ at small
positive arguments $0 \!<\! \D/G_{0}^{\scriptscriptstyle (4)} \!\ll\! 1$ and
decays exponentially with $p(\D) \approx p(0) (4G_{0}^{\scriptscriptstyle (4)}
/3\pi \D)^{1/2} \exp(-\D/4G_{0}^{\scriptscriptstyle (4)})$ for large values 
$\D/G_{0}^{\scriptscriptstyle (4)} \gg 1$.

With the full expression for $p(\D)$ at hand, the stable area fraction
$\mathcal{C}_{\mathrm{pos}}$ of the two-dimensional (Gaussian-distributed)
potential landscape can be determined: It is convenient to use the expression
\eqref{eq:probability-distribution-Dlarger0} and integrate over $\D$ first;
the subsequent integral over $q$ then yields the universal result
\begin{align}\label{eq:Cpos-Gaussian}
   \mathcal{C}_{\mathrm{pos}} = (3-\sqrt{3})/6 \approx 0.21,
\end{align}
independent of $G_{0}^{\scriptscriptstyle (4)}$ and thus of the shape of the
correlator. We find that for a Gaussian random potential the stable area
involves about one-fifth of the total landscape; in physical terms
it means that only a small fraction the landscape can be explored by pinscape
spectroscopy, while a large portion (nearly 80\%) of the plane are either
unstable or indefinite areas.

\subsection{Intermediate defect densities}\label{sec:intermediate}

At intermediate densities, we have to resort to numerical studies; these will
provide us---besides the desired information on the stable fraction
$\mathcal{C}_{\mathrm{pos}}$---with some additional insights on the fraction
of unstable ($\mathcal{C}_{\mathrm{neg}}$) and indefinite regions
($\mathcal{C}_{\mathrm{ind}}$) of such random landscapes.

\begin{figure}[t]
\centering
\includegraphics[width = .47\textwidth]{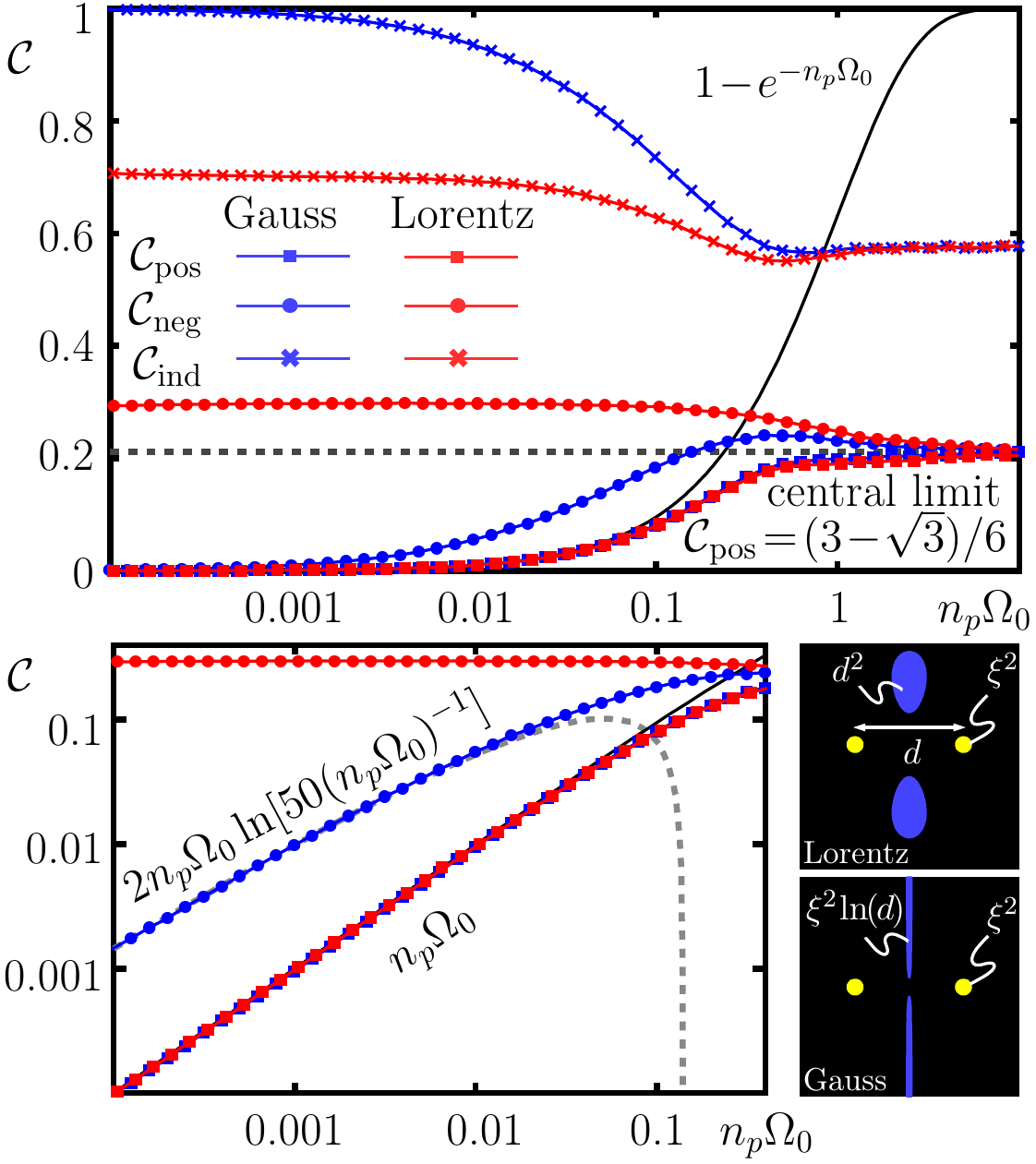}
\caption{Fraction of stable ($\mathcal{C}_\mathrm{pos}$, squares), unstable
($\mathcal{C}_\mathrm{neg}$, circles), and indefinite
($\mathcal{C}_\mathrm{ind}$, crosses) areas of a potential landscape
characterized by a finite density $n_p$ of Lorentzian
[$V_{\mathrm{\scriptscriptstyle L}}(r)$, red] or Gaussian
[$V_{\mathrm{\scriptscriptstyle G}}(r)$, blue] shaped defects potentials,
respectively.  The log-linear scale (top) highlights the behavior at large
densities, while the scaling at low densities is more prominent in the log-log
representation (bottom).  At small densities the fraction of stable points
follows the universal law $\mathcal{C}_{\mathrm{pos}} = n_{p} \Omega_{0}$, see
bottom figure. At large densities the Hessian character approaches that of a
random potential with Gaussian correlator (black dashed line in top
panel).  The stable area fraction of the cut parabolic trap
$V_{\mathrm{\scriptscriptstyle L}}(r)$ is shown as a black line in the
top panel.  At low defect densities $n_p\Omega_0 \!\to\! 0$, see bottom
panel, the unstable fraction $\mathcal{C}_\mathrm{neg}$ reaches a constant
value for the Lorentzian-shaped potential (red circles) and decays as
$\mathcal{C}_\mathrm{neg} \sim n_p\Omega_0 \ln[(n_p\Omega_0)^{-1}]$ for the
Gaussian-shaped potential (blue circles).  This is owed to the different
scaling of unstable regions defined by distant defects in the dilute limit, as
shown in the two thumbnails on the bottom right with yellow (stable), blue
(unstable/maxima), and black (unstable/saddle points) areas.
}
\label{fig:Gauss-Lorentz-numerical}
\end{figure}

We have explored this regime for the three different types of defect
potentials, cut parabolas $V_{\mathrm{\scriptscriptstyle P}}(r)$,
Lorentzian-shaped $V_{\mathrm{\scriptscriptstyle L}}(r)$ with algebraic tails,
and short-range Gaussian-shaped $V_{\mathrm{\scriptscriptstyle G}}(r)$, and
computed the area fractions $\mathcal{C}_{\mathrm{pos}}$,
$\mathcal{C}_{\mathrm{neg}}$, and $\mathcal{C}_{\mathrm{ind}}$ for stable,
negative-definite and indefinite regions, respectively. This numerical
analysis reveals several interesting facts, see
Fig.~\ref{fig:Gauss-Lorentz-numerical}: First, the Hessian character
$\mathcal{C}_{\mathrm{pos}}$ grows linearly from zero (at low densities).
For the regular potentials $V_{\mathrm{\scriptscriptstyle L}}(r)$ and
$V_{\mathrm{\scriptscriptstyle G}}$, the stable fraction saturates rapidly
(i.e., for $n_{p}\Omega_{0} \gtrsim 4$) to the value obtained for a Gaussian
random pinscape, with the precise functional dependence on the density
parameter $n_{p}\Omega_0$ differing numerically. The (irregular) cut parabolas
$V_{\mathrm{\scriptscriptstyle P}}(r)$, however, behave differently, with the
entire area becoming stable at large densities $n_p$,
$\mathcal{C}_{\mathrm{pos}} \to 1$, see below for more details.

A qualitative difference is observed between $V_{\mathrm{\scriptscriptstyle
L}}$ and $V_{\mathrm{\scriptscriptstyle G}}$ for the negative-definite
area-fraction $\mathcal{C}_{\mathrm{neg}}$ (the latter vanishes for
$V_{\mathrm{\scriptscriptstyle P}}$). This quantity assumes
a macroscopic value $\sim 30\%$ for the long-range Lorentzian traps, while
vanishing at low densities for the Gaussian-shaped pins, see
Fig.~\ref{fig:Gauss-Lorentz-numerical} (bottom).
The difference is attributed to the long-range, i.e.\ power-law, nature of
the potential
and can be understood by considering a pair of defects: For a single
(rotationally symmetric) defect, the transverse curvature (along the azimuth)
is always positive, while the longitudinal curvature (along the radius)
changes from positive near the center to negative further out. Hence, a single
defect generates either minima or saddles and a pair of defects is required to
produce a maximum through proper superposition of the two negative
longitudinal curvatures.

For a pair of defects with long-ranged potential (e.g., Lorentzian) at a distance $d$, the decay of the tails [$V(r) \!\sim\! r^{-\alpha}$, $\alpha \!>\! 1$] has no intrinsic length scale, and the area of regions with negative curvature scales as $d^{2}$.
This area becomes anisotropic [$\mathrm{width} \!\times\! \mathrm{height}
\!\approx\! (d/\sqrt{\alpha})\!\times\!(\sqrt{\alpha} d)$, see thumbnail in
Fig.~\ref{fig:Gauss-Lorentz-numerical}] as $\alpha$ increases.
At low defect density, the height $\sqrt{\alpha} d$ gets cut off by the
typical inter-defect distance $d \!=\! n_{p}^{-1/2}$, resulting in a concave
area $\propto d^2 / \sqrt{\alpha}$. The area fraction 
$\mathcal{C}_{\mathrm{neg}} \!\propto\! (n_p\Omega_0)^0 / \sqrt{\alpha}$ with negative curvature is non-vanishing in the limit $n_{p} \!\to\! 0$.
For short-ranged defects, i.e., where a length-scale $\xi$ dictates the decay away from the defect, the result is not universal as it depends on the negatively curved overlap
produced by two distant defects.
Specifically, for two defects separated by $d$, the negative
overlap is limited to a
slim area concentrated near the normal (line) to the midpoint between the
defects (Wigner-Seitz or Voronoi decomposition), see thumbnails in
Fig.~\ref{fig:Gauss-Lorentz-numerical}.
For Gaussian-shaped defect potentials, the area fraction can be
evaluated to $(\xi/d)^{2} \ln(d/\xi)$, yielding $\mathcal{C}_{\mathrm{neg}}
\propto n_{p}\Omega_0 \ln[(n_{p}\Omega_0)^{-1}]$.

The special case of cut parabolas $V_{\mathrm{\scriptscriptstyle P}}(r)$ can
be treated analytically, since curvatures are non-negative integer multiples
of $2 V_0/\xi^2$.  More specifically, within a defect's range of action $r <
\xi$, the Hessian matrix $H = (2 V_0/\xi^2) \mathbb{I}$ is position
independent, diagonal, and positive definite, while it vanishes outside.  As a
result, non-overlapping traps act as isolated ones, while the total Hessian
determinant of $\nu$ overlapping traps is $\nu^{2} (2V_0/\xi^2)^{2} \geq 0$.
We thus conclude that the only non-stable (and hence indefinite) regions are
those where no defect is active, i.e., where $\nu = 0$. This probability is
given by the zeroth term of the Poisson distribution $\Poiss(\nu,
n_{p}\Omega_0) = (n_{p}\Omega_0)^{\nu}\exp(-n_{p}\Omega_0)/\nu!$ (see Appendix
\ref{app:parabola} for a detailed discussion) and hence the complement defines
the stable area,
\begin{align}\label{eq:Poisson-result}
   \mathcal{C}_{\mathrm{pos}} = 1 - \Poiss(0,n_{p}\Omega_{0}).
\end{align}
This area fraction approaches unity at large defect densities $n_p$, see black
line in Fig.~\ref{fig:Gauss-Lorentz-numerical} and top panel in
Fig.~\ref{fig:stability-map-highdensity}, quite different from the other two
examples of Gaussian and Lorentzian shaped potentials that approach the
Gaussian limit $\mathcal{C}_{\mathrm{pos}} \approx 21\%$. This is due to the
singular property of the cut parabola that does not provide any region with a
negative definite Hessian; when the parabolas are cut rather than smoothly
connected to zero, only convex and flat regions appear in the pinning
potential landscape.

\begin{figure}[t]
\centering
\includegraphics[width = .4\textwidth]{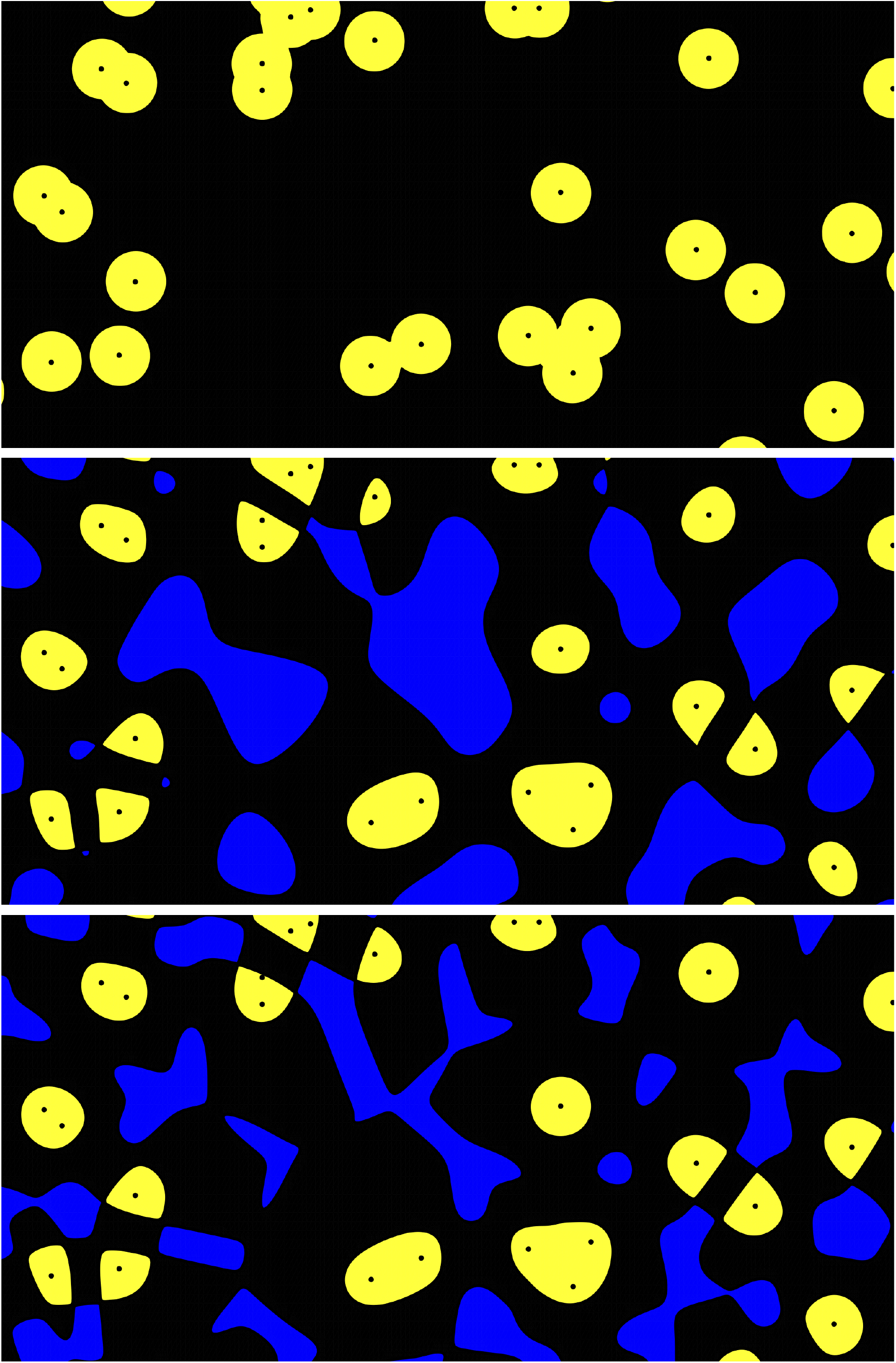}
\caption{Hessian map of a potential landscape for a moderate density of cut
parabolic (top), Lorentzian (middle), and Gaussian (bottom) traps. To allow
for direct comparison, the defect position is the same in all panels (we chose
a density parameter $n_p \xi_{\scriptscriptstyle \mathrm{G}}^{2} \!=\! 0.125$
and the view area $100\, \xi_{\scriptscriptstyle \mathrm{G}}\!\times\!
50\,\xi_{\scriptscriptstyle \mathrm{G}}$) and the length $\xi$ (defining the
defect shape) assumes the values $(1/2)^{1/2}\xi_{\scriptscriptstyle
\mathrm{G}}$, $(3/2)^{1/2}\xi_{\scriptscriptstyle \mathrm{G}}$, and
$\xi_{\scriptscriptstyle \mathrm{G}}$ respectively. This implies an elementary
area fraction of $n_p \Omega_0 \approx 0.2$ for all three cases. Yellow/blue
denote stable/unstable regions where the Hessian matrix is positive/negative
definite. Indefinite points are colored in black. Here, the difference in the
area fraction $\mathcal{C}_\mathrm{neg}$ of unstable points for the Lorentzian
($\sim 28\%$) and Gaussian ($\sim 22\%$) traps is apparent, see
Fig.~\ref{fig:Gauss-Lorentz-numerical}.}
\label{fig:stability-map}
\end{figure}
\begin{figure}[t]
\centering
\includegraphics[width = .4\textwidth]{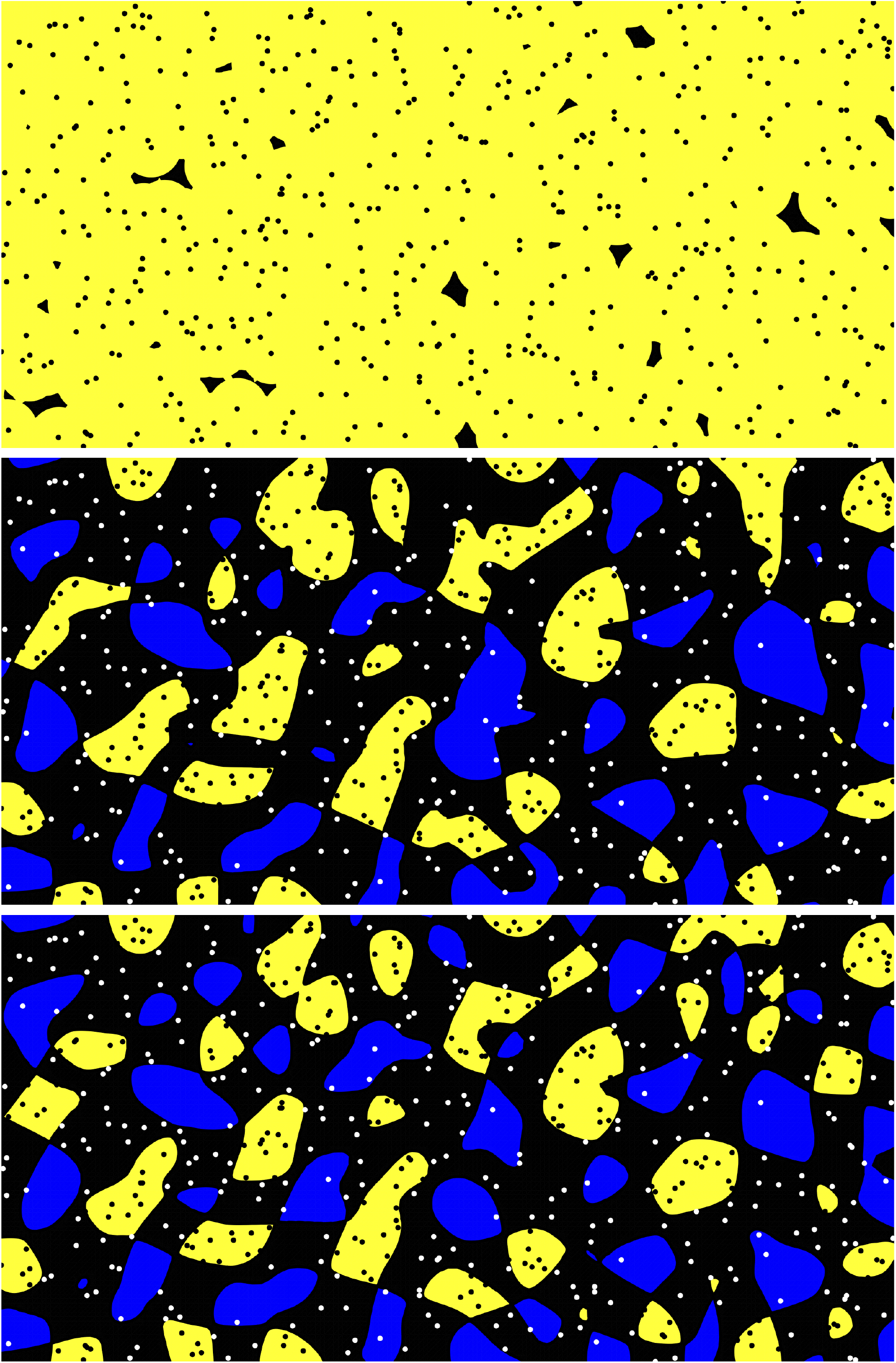}
\caption{Hessian map of a potential landscape for a high density of cut
parabolas (top), Lorentzian (middle), and Gaussian (bottom) traps. The defect
position is equal in all panels (we chose a density parameter $n_p
\xi_{\scriptscriptstyle \mathrm{G}}^{2} \!=\!  2.5$ and a view area
$100\,\xi_{\scriptscriptstyle \mathrm{G}}\!\times\!
50\,\xi_{\scriptscriptstyle \mathrm{G}}$). The length parameter $\xi$
(defining the defect shape) assumes the values
$(1/2)^{1/2}\xi_{\scriptscriptstyle \mathrm{G}}$,
$(3/2)^{1/2}\xi_{\scriptscriptstyle \mathrm{G}}$, and $\xi_{\scriptscriptstyle
\mathrm{G}}$ respectively. The elementary area fraction is $n_p \Omega_0
\approx 4$. Yellow/blue denotes stable/unstable regions where the Hessian
matrix is positive/negative definite. Indefinite points are black.  Dense
defect clusters (black points in yellow domains) define stable pinning
regions, low density areas (white defects in blue regions) are unstable.}
\label{fig:stability-map-highdensity}
\end{figure}
\begin{figure*}[tbh]
\centering
\includegraphics[width = \linewidth]{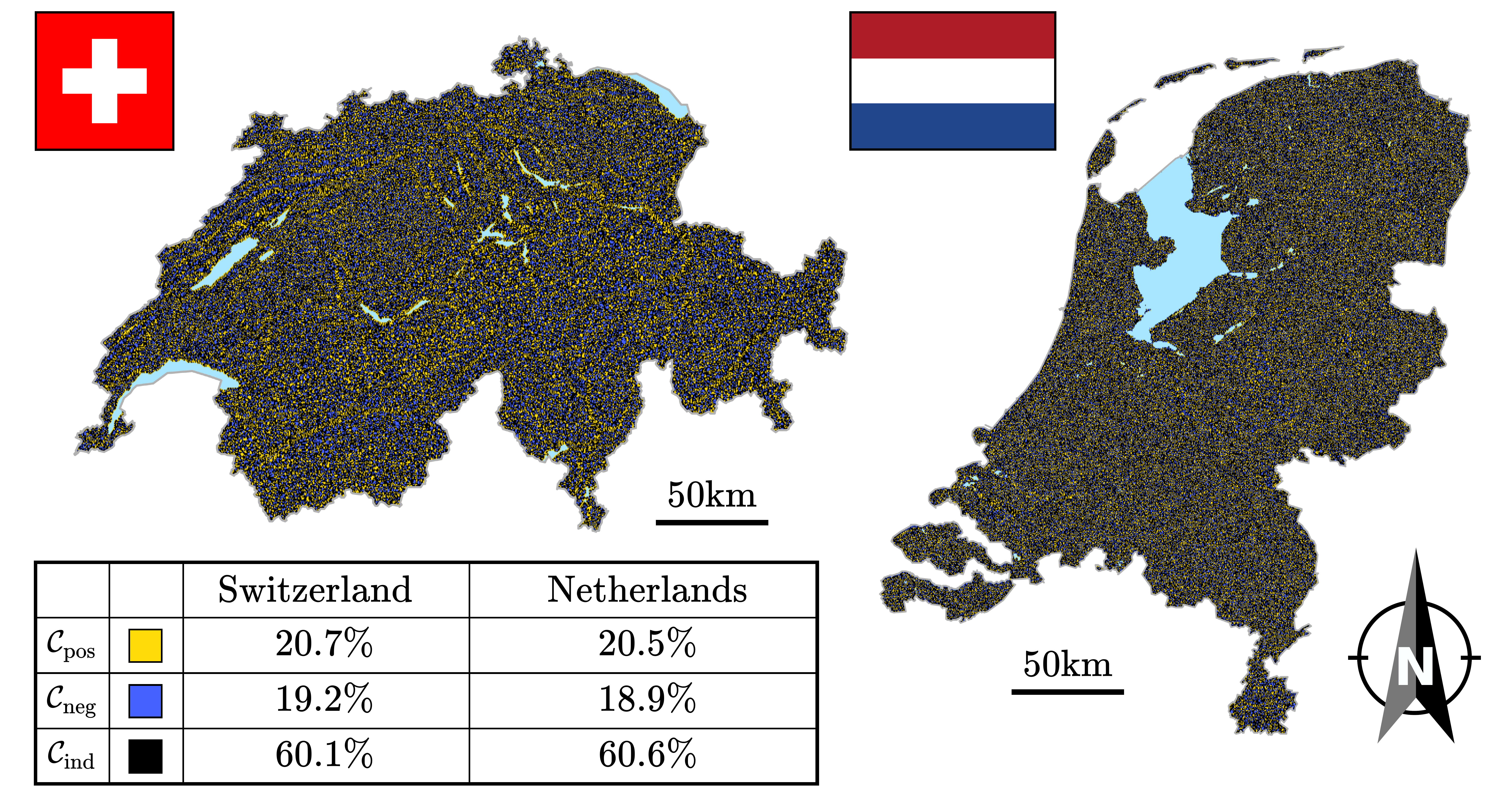}
\caption{
Hessian Map of two natural landscapes: For Switzerland (left) and the
Netherlands (right) stable, unstable, and indefinite areas are colored in
yellow, blue, and black respectively.
From an elevation map on a square lattice with longitude and latitude
angular resolution of %$1/240^{\circ}\mathrm{deg}$
$15~\mathrm{arcseconds}$ [data from Wolfram Mathematica's geographic data package], the Hessian matrix is evaluated by fitting a quadratic polynomial through each $3 \!\times\! 3$ plaquette.
Despite the two countries having very different topography, their Hessian characters---tabled above---are close to the Gaussian result $\mathcal{C}_{\mathrm{pos}} = \mathcal{C}_{\mathrm{neg}} \approx 21\%$, see Eq.~\eqref{eq:Cpos-Gaussian}. We thus surmise a universal Hessian law for natural landscapes.
}
\label{fig:landscapes}

\end{figure*}

Figures \ref{fig:stability-map} and \ref{fig:stability-map-highdensity}
illustrate our findings for the two cases of low, $n_p\Omega_0 = 1/5$, and
high density parameters $n_p\Omega_0 = 4$, respectively.  In Figure
\ref{fig:stability-map}, we show the Hessian map for a moderate density of cut
parabolic (top), Lorentzian (middle), and Gaussian (bottom) defects.  For the
cut parabolas, the Hessian determinant assumes only discrete values that
follow from the number of overlapping defects. While the shape of stable
regions (yellow) are trivial for the cut parabolas, this is no longer the case
for the Lorentzian/Gaussian potentials. In Figure
\ref{fig:stability-map-highdensity}, we show the Hessian map for a large
density of cut parabolic (top), Lorentzian (middle), and Gaussian (bottom)
defects.  For the cut parabolas, the Hessian determinant guarantees stability
in almost every point on the map. For the smooth Lorentzian and Gaussian
potentials, different pins mutually neutralize one another and the stable
regions are more scarce.  Only when defects cluster, they reinforce one
another to produce stable regions, see black dots in yellow regions. On the
contrary, dilute regions with fewer defects than average (white dots in blue
domains) produce unstable regions.

\section{Summary and Conclusion}\label{sec:summary}

Inspired by the recent advances in vortex imaging and the development of
pinscape spectroscopy, we have analyzed the properties of 2D pinning
landscapes with the help of a new characteristics, the Hessian matrix
$H(\vec{r})$, its determinant $\det H$, and its trace $\tr H$.  We have
introduced the \emph{Hessian stability map} as a bi-colored map that separates
stable from unstable regions of the pinscape; while stable regions can be
mapped via pinscape spectroscopy using appropriate (linear) driving forces,
unstable regions cannot, i.e., these regions do not provide equilibrated
vortex positions for any applied (linear) force. We have drawn attention to
several peculiarities of pinscape spectroscopy (the so-called
`sping-softening' and `broken spring effects' in
Ref.~[\onlinecite{Embon2015}]) related to the stability boundaries of the
Hessian map where the determinant $\det H$ vanishes, e.g., an enhanced
response involving potential non-linearities as well as the thermal activation
over barriers into the unstable regions.  Furthermore, we have indicated how
pinscape spectroscopy can be enhanced to cover extended regions around the
vortex trajectories by probing the out-of-phase response of vortices at high
frequencies.

Second, we have introduced the \emph{Hessian character}
$\mathcal{C}_\mathrm{pos}$ of a pinning landscape $U(\vec{r})$ as the area
fraction of the plane that covers the stable regions of the Hessian map. We
have investigated two types of generic pinscapes, those arising from a random
distribution of defects with individual pinning potentials $V(\vec{r})$ and
the case of a Gaussian random potential characterized through its correlator
$G(\vec{r})$. Different individual defect potentials $V(\vec{r})$ have been
studied, cut parabolas with a discrete Hessian map and an exceptionally large
stable fraction $\mathcal{C}_\mathrm{pos} \to 1$ at large defect densities
$n_p\Omega_0 \gg 1$, Lorentzian-shaped trapping potentials that induce
correlations through their long-range tails and produce a finite unstable
fraction $\mathcal{C}_\mathrm{neg}$ in the limit of small defect density
$n_p\Omega_0 \ll 1$, and Gaussian shaped potentials with a short range that
behave most regularly at all densities.  The Hessian character of both,
Gaussian and Lorentzian potentials, approaches the character of the random
Gaussian potential for large defect densities $n_p \Omega_0 \gg 1$, with the
latter assuming a universal value of $\mathcal{C}_\mathrm{pos} \approx
21\,\%$ independent of the correlator $G(\vec{r})$.  Hence, we find that
pinscape spectroscopy of regular pinning potentials can probe at most a
fraction of about one-fifth of the plane.

Unfortunately, up to now, the 'vortex in the maze' experiment is limited to a
single tunable drive parameter. This is owed to the experimental setup
measuring the vortex motion in the region of a current-driven strip. An
expanded view on the pinscape within this setup can be gained by injecting the
vortex at different positions along the transverse ($y$) direction.  However,
other geometries allowing for different drive directions may open the
possibility to probe the full stable region of a pinscape, thus coming closer
to the original 'ball-in-the-maze' setup also for the vortex.  

Finally, the Hessian of pinning potentials $U(\vec{r})$ turns out
relevant in the discussion of strong pinning physics \cite{Buchacek2020}, see
also Refs.~\cite{Tanguya2004, Cao2018}, specifically near the onset of strong
pinning as described by the famous Labusch criterion \cite{Labusch1969}:
Within the strong pinning paradigm, the many body problem of vortex lattice
pinning is reduced to the minimization of the two-dimensional total pinning
energy $e_\mathrm{pin}(\vec{r}) = \bar{C} (\vec{r}-\vec{x})^2/2 + V(\vec{r})$
including both an elastic energy (with $\bar{C}$ an effective elastic
constant) and $V(\vec{r})$ the pinning potential of an individual defect.
Under strong pinning conditions with $V(\vec{r})$ dominating the elastic term,
the position $\vec{r}$ of the pinned vortex undergoes pinning and depinning
jumps as the lattice moves smoothly along $\vec{x}$, similar to our vortex in
the plane that gets trapped and detrapped by stable regions of the pinscape.
Indeed, expanding the total pinning energy $e_\mathrm{pin}(\vec{r}) =
\bar{C}\,x^2/2 - \bar{C}\, \vec{r}\cdot \vec{x} + V_\mathrm{eff}(\vec{r})$
with the renormalized effective potential $V_\mathrm{eff}(\vec{r}) =
V(\vec{r}) + \bar{C}\,r^2/2$ (the term $\bar{C}\, x^2/2$ is an irrelevant
shift), we reduce the strong pinning problem to the vortex-in-the-maze problem
with the elastic term $\bar{C}\, \vec{r}\cdot\vec{x}$ replacing the external
drive $\vec{F}_\mathrm{\scriptscriptstyle L} \cdot \vec{r}$ due to the
current-induced Lorentz force (incidentally, the lattice coordinate $\vec{x}$
is driven by the applied current density $\vec{j}$ as well).  This equivalence
opens up interesting new avenues in the strong pinning problem
\cite{Buchacek2020}.

Besides this relation to strong pinning, one might think of completely
different applications of Hessian maps and characters, a quite obvious one
that comes to mind are natural (topographic) landscapes. Indeed,
analyzing the elevation map of different topographic landscapes---we chose Switzerland and the Netherlands as examples, see Fig.~\ref{fig:landscapes}---one finds in both cases the
characters $\mathcal{C}_\mathrm{pos} \!\approx\! 21\,\%$, $\mathcal{C}_\mathrm{neg} \!\approx\! 19\,\%$ and $\mathcal{C}_\mathrm{ind} \!\approx\! 60\,\%$, close to the value for the Gaussian random landscape. This raises interesting questions about universality and the (non-)Gaussianity of
natural landscapes.

\begin{acknowledgments}
We wish to express our special thanks to Eli Zeldov for initiating and
supporting this project, to Yonathan Anahory for providing experimental input,
and to Gian Michele Graf, who helped us formulating and solving the path
integral problem of the Hessian matrix. The authors acknowledge financial
support of the Swiss National Science Foundation (SNSF) through the NCCR
MaNEP. R.W.\ further acknowledges the support from the Pauli Center for
Theoretical Studies at ETH Zurich through its scientific visitor program and
the Heidelberger Akademie der Wissenschaften (WIN, 8.\ Teilprogramm).
\end{acknowledgments}

\appendix

\section{Parameter $\eta\omega/U''$}\label{app:parameter}

We derive an estimate for the parameter $\eta\omega/U''$ governing the
response $\vec{u}$. Typical values for this ratio are obtained from the
estimate $E_{\mathrm{pin}} \sim (H_{c}^{2}/4\pi)\, \xi^{2}d_{s}$ of the vortex
core energy in a film of thickness $d_{s}$; here, $H_c = 
\Phi_{0}/2\sqrt{2}\pi \lambda \xi$ denotes the thermodynamic critical field
and $\lambda$ and $\xi$ are the penetration depth and the coherence length,
respectively. The coherence length provides an estimate for the typical
spatial variation in the pinscape and hence $U'' \sim E_{\mathrm{pin}} /
\xi^{2}$. The viscosity $\eta$ follows from the Bardeen-Stephen
 \cite{Bardeen1965} formula $\eta = \Phi_{0}^{2} d_{s}/ 2\pi \xi^{2} \rho_{n}
c^{2}$, with the flux quantum $\Phi_{0} = hc/2e = 2.07 \times
10^{-7}~\mathrm{Gcm^{2}}$. Inserting the Drude expression $\rho_n = m / n
e^{2}\tau$ for the normal state resistivity, where $n$ is the electronic
density and $\tau$ the electron relaxation (scattering) time, we find the
ratio
\begin{align}\label{}
   \eta \omega / U'' \sim (n/n_{s}) \omega \tau
\end{align}
with $n_s$ the superfluid density. Assuming a value $n/n_s$ of order unity, we
find the parameter $\eta \omega/U''$ to be small in general. E.g., in the
experiment on Pb-films of Ref.~[\onlinecite{Embon2015}], the parameters $\xi =
46~\mathrm{nm}$, $\lambda \approx 90~\mathrm{nm}$, and $d_{s} =
75~\mathrm{nm}$ provide an estimate $E_{\mathrm{pin}}/ \xi^2 \approx
7.5\times10^{-5}~\mathrm{N/m}$. Assuming a normal state resistivity $\rho_{n}
\approx 0.01~\mathrm{\mu\Omega cm}$ for lead \cite{Montgomery1958}, we find
that $\eta \approx 2.4\times10^{-13}~\mathrm{N s / m}$ and combining this
estimate with the $ac$ frequency $\omega = 13.3~\mathrm{kHz}$ of the
experiment, we arrive at $\eta\omega \approx 3.2 \times 10^{-9}~\mathrm{N/m}$,
a value that is 3--4 orders of magnitude lower than typical curvatures $U''$.

\section{2D local reconstruction of pinscape}\label{app:1st-order}

The solution of the equation of motion \eqref{eq:eom} provides us with the
expressions
\begin{align}\label{eq:general-ux}
   \!\!\frac{u_{x}}{F_{ac}} &\!=\! \frac%
   {4b^{2} + \eta^{2}\omega^{2}}%
   {2b (4ab \!-\! c^{2}) + 2a \eta^{2}\omega^{2}
   + i \eta \omega[4b^{2} \!+\! c^{2} \!+\! \eta^{2}\omega^{2}] },\!\!\\
   \label{eq:general-uy}
   \!\!\frac{u_{y}}{F_{ac}} &\!=\! \frac%
   {-c (2b - i\eta\omega)}%
   {2b (4ab \!-\! c^{2}) + 2a \eta^{2}\omega^{2}
   + i \eta \omega[4b^{2} \!+\! c^{2} \!+\! \eta^{2}\omega^{2}] }\!\!
\end{align}
for the displacements $u_x$ and $u_y$. This result can be analyzed
perturbatively in the small parameter $\eta\omega/U''$ and leads us to the
simple expression Eq.~\eqref{eq:zeroth-order-solution} to lowest (0-th) order.
The expansion of Eqs.~\eqref{eq:general-ux} and \eqref{eq:general-uy} to
linear order in $\eta \omega/U''$ contributes the out-of-phase displacements
$\delta u_x, ~\delta u_y \propto i(\eta\omega/U'')(F_{ac}/U'')$ that
allow for the full local construction of the pinscape $U(x,y)$ in the vicinity
of the vortex trajectory. Specifically, this out-of-phase response assumes the
form
\begin{align}\label{eq:first-order-solution}
   \!\!\!\!
   \frac{\delta u_{x}}{F_{ac}} &= - i \eta \omega \frac{4b^{2} + c^{2}}
   {(4a b - c^{2})^{2}}, &\;\;
   \frac{\delta u_{y}}{F_{ac}} &= i \eta \omega \frac{2(a+b)c}{(4a b - c^{2})^{2}}
\end{align}
and can be measured independently from the in-phase displacements in Eq.~\eqref{eq:zeroth-order-solution}.  For a fixed drive amplitude $F_{ac}$, the
independent measurement of the four quantities $u_{x}$, $u_{y}$, $\delta
u_{x}$, and $\delta u_{y}$ then allows to extract all the local curvatures
$a$, $b$, and $c$ from the experiment,
\begin{align}\label{eq:a}
   a &= \frac{F_{ac}}{2u_x} \Big[1
   +\frac{u_y^2/u^2}{(\delta u_{y}/u_y)(u_x/\delta u_{x})-1}\Big],\\
   \label{eq:b}
   b &= \frac{F_{ac}}{2u_{x}} \frac{u_x^2/u^2}{(\delta u_{y}/u_y)(u_x/\delta u_{x})-1},\\
   \label{eq:c}
   c &= \frac{F_{ac}}{2u_{x}} \frac{-2u_x u_y/u^2}{(\delta u_{y}/u_y)(u_x/\delta u_{x})-1},
\end{align}
where $u = (u_{x}^{2} + u_{y}^{2})^{1/2}$ is the total displacement amplitude.
The additional independent relation $\eta \omega = F_{ac}\,|\delta u_{x}|
/u^{2}$ with a constant left-hand side $\eta \omega$ serves as a check.  The
results \eqref{eq:a}--\eqref{eq:c} can be used to reconstruct the potential in
the vicinity of the trajectory.  We define the vector $\vec{\eta}_{\perp}
\equiv (1, -u_x/u_y) = (1, 2b_{n}/ c_{n})$ perpendicular to the vortex
trajectory and parametrize the positions $\vec{r}_{n,\epsilon} = \vec{r}_{n} +
\epsilon \vec{\eta}_{\perp}$ transverse to the equilibrium trajectory at
$\vec{r}_{n}$.  Combining Eqs.~\eqref{eq:a}--\eqref{eq:c} and
\eqref{eq:tilt-again}, we find the potential shift
\begin{align}\label{eq:U-transverse}
\!\!
   U(\vec{r}_{n,\epsilon}) &\!-\! U(\vec{r}_{n})
   = \epsilon F_{\mathrm{\scriptscriptstyle L}n} + \epsilon^{2} [a_{n} + 2 b_{n}
     + 4b_{n}^{3}/c_{n}^{2}]\\ \nonumber
     &= \epsilon F_{\mathrm{\scriptscriptstyle L}n} + \epsilon^{2} \frac{F_{ac}}{2 u_x}
     \Big[ 1 + \frac{u^2/u_y^2}{(\delta u_{y}/u_y)(u_x/\delta u_{x})-1}\Big].
\end{align}
While the linear term $\propto \epsilon$ in the bare potential is 'tilted
away' by the force $F_{\mathrm{\scriptscriptstyle L}n} = n F_{ac}$, the quadratic term $\propto
\epsilon^{2}$ provides the parabolic confinement transverse to the vortex
trajectory.
Unfortunately, the corrections Eq.~\eqref{eq:first-order-solution} are small
in the parameter $\eta\omega/U''$, requiring a high measurement sensitivity
and $ac$ frequencies in the MHz range.

The solutions Eqs.~\eqref{eq:zeroth-order-solution} and
\eqref{eq:first-order-solution} for the in-phase and out-of-phase motion apply
when $\delta u_{x}/u_{x},\ \delta u_{y}/u_{y} \ll 1$, i.e., when
\begin{align}\label{eq:lin-exp-crit}
   \eta \omega \ll \frac{2b (4ab - c^{2})}{4b^{2}+c^{2}}
   \quad \text{and} \quad
   \eta \omega \ll \frac{4ab - c^{2}}{2(a + b)}.
\end{align}
These criteria are violated in the vicinity of the \emph{Hessian boundary}
where the condition $4ab - c^2 = 0$ is separating a stable from an unstable
region. Near this boundary, the singularities in Eq.~\eqref{eq:zeroth-order-solution} are cut off by the dissipative term
$\eta\omega$ and the appropriate solutions to linear order in
$F_{ac}/\eta\omega$ take the form
\begin{align}\label{eq:intermediate-solution}
   \frac{u_{x}}{F_{ac}} &= \frac{-i}{\eta\omega}
   \frac{1}{1\!+\!(c/2b)^{2}},\quad\quad \frac{u_{y}}{F_{ac}} =
   \frac{i}{\eta\omega} \frac{c/2b}{1\!+\!(c/2b)^{2}}.
\end{align}
These displacements are phase-lagged with respect to the external drive, while
the motion is still at the same angle $\phi$ away from the $x$ axis.

\section{Escape}\label{app:escape}

Here, we comment on the escape of the vortex from the stable region when
approaching the Hessian boundary. The quadratic approximation
\eqref{eq:tilt-potential} then is insufficient to describe the escape dynamics
over the depinning barrier. The latter is obtained by including cubic terms in
the expansion; limiting ourselves to the most relevant term $d\, u_x^3$,
we obtain the expansion around the position $\vec{r}_0$ near the boundary
\begin{align}\nonumber
   U_{\mathrm{tilt}}(\vec{r},F_{\mathrm{\scriptscriptstyle L}})
   = U_{\mathrm{tilt}}(\vec{r}_{0},F_{\mathrm{\scriptscriptstyle L}})
   + a u_x^{2} + b u_y^2
   + c u_x u_y + d u_x^3
\end{align}
with $d<0$ describing the escape for positive tilt. This potential features a
saddle point at
\begin{align}\label{eq:saddle-point-position}
   \vec{r} = \vec{r}_0-\frac{2\tilde{a}}{3d} (1, -c/2b)
\end{align}
and defines a barrier
\begin{align}\label{eq:dc-barrier}
   U_{b} = 4\tilde{a}^3 / 27d^{2}
\end{align}
that prevents the escape of the vortex to the unstable region; here, we have
introduced the renormalized curvature $\tilde{a} = a (1 - c^{2}/4ab)$, which
scales linearly with the Hessian determinant and vanishes upon approaching the
stability edge. Note that the curvature parameters in the above expressions
depend on $\vec{r}_0$ and hence on the closeness of this point to the Hessian
stability boundary.

At finite temperature, the vortex escapes the defect by thermal activation
when the criterion $U_b \approx k_{\rm\scriptscriptstyle B} T \ln(\omega_0
\tau)$ is met, with $\omega_0$ the attempt frequency for escaping the well and
$\tau$ the relevant time scale of the experiment \cite{Kramers1940,
Haenggi1990}. In order to better understand the situation in the experiment of
Ref.~\cite{Embon2015}, we can use these relations to find the
distance $\delta r = |\vec{r} - \vec{r}_{0}|$ away from the boundary where the
vortex leaves the pin via thermal activation. Using the
estimates \cite{Embon2015} $\omega_0 \sim 10^{11}$ Hz and $\tau \sim
300~\mathrm{s}$, we find that $U_b \approx 30\, k_{\rm\scriptscriptstyle B} T
\approx 130$ K at the temperature $T = 4.2$ K of the experiment. Combining the
expressions for the saddle point position \eqref{eq:saddle-point-position},
for the barrier \eqref{eq:dc-barrier} and for the displacement $u_x =
F_{ac}/2\tilde{a}$, see \eqref{eq:zeroth-order-solution}, we obtain
\begin{align}\label{}
   \delta r \approx [(u_x^2 + u_y^2) (6U_b/F_{ac} u_x) ]^{1/2}.
\end{align}
For the escape out of the well at $x \!\approx\! 20$ nm (right edge of the
central well in Fig.~\ref{fig:stability-region}), where $(u_x, u_y)
\!\approx\!  (0.15, -0.05)$ nm and with $F_{ac} \!\approx\! 10^{-14}$ N, one
arrives at a typical energy change per step in $F_\mathrm{\scriptscriptstyle
L}$ of $F_{ac} u_x \!\approx\! 0.1$ K.  This results in an estimate $\delta r
\!\approx\! 14$ nm, an appreciable distance away from the Hessian stability
boundary.  Hence, one has to conclude that thermal fluctuations cut off the
measured trajectory long before reaching the Hessian stability boundary, in
agreement with the discussion in the experiment \cite{Embon2015}. As a
consequence, the displacements $u_x$ and $u_y$, although proportional to the
inverse Hessian $(4ab - c^2)^{-1}$, do not show a divergence when approaching
the Hessian stability boundary, as the latter is never closely approached. In
the same vain, the vortex leaves the pin much before the force saturates at
the Hessian boundary.

In principle, anharmonic effects may influence the vortex escape from the
stable regions---this is the case at small temperatures [when $U_b  \gg \kB T
\ln(\omega_0 \tau)$] or at large $ac$ amplitudes $u$. Including such
anharmonicities and solving for the displacement $u_x$, we find the periodic
dynamics
\begin{align}\label{eq:cubic-solution}
   u_{x} = -\frac{2b}{c}u_{y} &=
   \frac{\tilde{a}}{3d} \Bigg[\sqrt{1 + \frac{3 d}{ \tilde{a}^{2}}
   F_{ac}e^{-i\omega t}} - 1\Bigg] e^{i\omega t},
\end{align}
as long as the $ac$ amplitude $F_{ac}$ is below the threshold
\begin{align}\label{eq:threshold}
   F_{\mathrm{thr}} \equiv \frac{\tilde{a}^{2}}{3 |d|}.
\end{align}
As the ratio $F_{ac}/F_\mathrm{thr}$ approaches unity, anharmonic effects
manifest; in particular, the barrier decreases periodically in time to a value
\begin{align}\label{eq:barrier}
   U_b^\mathrm{anh} = \frac{4 \tilde{a}^{3}}{27 d^{2}} \Big(1
   - \frac{F_{ac}}{F_{\mathrm{thr}}}\Big)^{3/2},
\end{align}
thus allowing for a faster escape of the vortex due to the combined effect of
thermal activation and anharmonicity in the $ac$ response. For even larger
$ac$ forces, $F_{ac} > F_{\mathrm{thr}}$, the vortex is pushed over the
barrier and leaves the defect for good. Expressing the ratio again through
known quantities, we find that $F_{ac}/F_{\mathrm{thr}} \approx \sqrt{F_{ac}
u_x/U_b}$ which, when inserting the experimental numbers \cite{Embon2015},
provides us with the value 1/20, telling us that anharmonic effects are small
for the experiment in Ref.~\onlinecite{Embon2015}.

\section{Gaussian probability distribution}\label{app:central-limit}

In the limit of strongly overlapping defects, the functional distribution
function $\mathcal{P} [U(\vec{r})]$ assumes a Gaussian form, see Eqs.~\eqref{eq:Gaussian-measure} and \eqref{eq:Gaussian-action}. We verify (and
sharpen) this statement by studying correlators and via direct calculation
of $\mathcal{P} [U(\vec{r})]$ from Eq.~\eqref{eq:DUr}.

\subsection{Correlators}\label{app:correlators}

Given a set of defect (or pin) locations $\{\vec{r}_{i}\}_{i=1}^N$,
we define the associated density
\begin{align}\label{eq:rho}
  \rho(\vec{r}) = \sum\nolimits_{i=1}^{N} \delta(\vec{r} - \vec{r}_{i}).
\end{align}
When distributed homogeneously over the area $\Omega$, the average density at
the position $\vec{r}$ is
\begin{align}\label{eq:rho_av}
  \langle \rho(\vec{r}) \rangle = \int \Big[\prod_{i=1}^{N}\frac{d^{2}r_{i}}{\Omega}\Big] 
  \rho(\vec{r}) = N/\Omega = n_p
\end{align}
and the two-point correlator reads
\begin{align}\label{eq:rhorho}
   \langle \rho(\vec{r}) \rho(\vec{s}) \rangle &=
      N(N \!-\! 1)/\Omega^2 + (N/\Omega)\> \delta(\vec{r}-\vec{s}).
\end{align}
Going over to reduced densities $\bar{\rho}(r) = \rho(r) -n_p$, the first four correlators
read (in the thermodynamic limit $N, \Omega \!\to\! \infty$, with $N/\Omega = n_p$)
\begin{align}\label{eq:rho^k}
   \langle \bar{\rho}(\vec{r}) \rangle &= 0,\\
       \nonumber
   \langle \bar{\rho}(\vec{r})\bar{\rho}(\vec{s}) \rangle &= n_p\, \delta(\vec{r}-\vec{s}),\\
       \nonumber
   \langle \bar{\rho}(\vec{r})\bar{\rho}(\vec{s})\bar{\rho}(\vec{t}) \rangle 
   &= n_p\, \delta(\vec{r}-\vec{s})\delta(\vec{r}-\vec{t}),\\
       \nonumber
   \!\!
   \langle \bar{\rho}(\vec{r})\bar{\rho}(\vec{s})\bar{\rho}(\vec{t})\bar{\rho}(\vec{x}) \rangle
   &= n_p\, \delta(\vec{r}-\vec{s})\delta(\vec{r}-\vec{t})\delta(\vec{r}-\vec{x}) \\
   \nonumber
   &\quad+ n_p^{2}\, \big[\delta(\vec{r}-\vec{s})\delta(\vec{t}-\vec{x})\\
   \nonumber
   &\quad +\! \delta(\vec{r} \!-\! \vec{t})\delta(\vec{s} \!-\! \vec{x})
   \!+\! \delta(\vec{r} \!-\! \vec{x}) \delta(\vec{s} \!-\! \vec{t})\big]\!.
\end{align}
These results translate into correlators for the potential
\begin{align}\label{eq:Ur-app}
   U(\vec{r}) = \sum\nolimits_i V(\vec{r}-\vec{r}_i) = \int d^{2}x \, V(\vec{r}-\vec{x}) 
   \rho(\vec{x})
\end{align}
via simple integration: $\langle U(\vec{r}) \rangle = 0$ (as $\int \!d^{2} r
V(\vec{r}) \!=\! 0$) and
\begin{align}\nonumber
   \langle U(\vec{r}) U(\vec{s}) \rangle &= n_p \! \int\! d^{2}x d^{2}y  
   V(\vec{r}-\vec{x}) V(\vec{s}-\vec{y})
   \langle \rho(\vec{x})\rho(\vec{y}) \rangle \\[-0.2em]
   &= G(\vec{r}-\vec{s}) \label{eq:UU}
\end{align}
with the two-point potential correlator
\begin{align}\label{eq:Green}
   G(\vec{r}-\vec{s}) =  n_p \xi^2 \int \frac{d^{2}x}{\xi^2} 
   V(\vec{r}-\vec{x}) V(\vec{s}-\vec{x}).
\end{align}
Here, $n_p\xi^2$ takes the role of the large density parameter, with the
integral remaining of order $V_0^2$. One easily shows that the even-order
$(2k)$-point correlators are dominated by the Wick term $\propto
(n_p\xi^2)^k$,
\begin{align}
   &\langle  U(\vec{r}_1) \cdots  U(\vec{r}_{2k}) \rangle =
   \\[-0.3em] \nonumber
   &\hspace{5em}
   \sum_{\substack{\mathrm{pairings}\\ \{p_{1},\dots, p_{k}\}}}\!\!\! 
   \Big[\prod_{\ell = 1}^{k}
   G(\vec{r}_{p_{\ell,1}} \!-\! \vec{r}_{p_{\ell,2}}) \Big]
   + \mathcal{O}[(n_p \xi^2)^{k-1}],
\end{align}
with the set of pairings $\{p_{1},\dots, p_{k}\}$ including all sites
$\vec{r}_{i}$ ($i \in \{1, \dots 2k\}$).
The odd-order ($2k+1$)-point correlators start with a subleading term
$\propto (n_p\xi^2)^k$. Note that all subleading terms involve higher-order
potential overlaps, e.g., the three-defect overlap of the form
\begin{align}\label{eq:Green-3}
\!\!\!
   G_3(\vec{r},\vec{s},\vec{t}) =  n_p \xi^2 \!\int\! \frac{d^{2}x}{\xi^2}
   V(\vec{r}-\vec{x}) V(\vec{s}-\vec{x}) V(\vec{t}-\vec{x}).\!\!
\end{align}
For large densities the Wick term dominates and the
distribution for $U(\vec{r})$ becomes Gaussian as $n_p\xi^2
\! \to\! \infty$.

\subsection{Probability distribution $\mathcal{P}[U(\vec{r})]$}

In order to calculate the functional probability distribution
$\mathcal{P}[U(\vec{r})]$, we discretize the problem and evaluate
$\mathcal{P}[\{U_\alpha\}]$ on the discrete set of lattice sites
$\{\vec{r}_\alpha\}_1^M$ on a mesh with unit volume $v = a^2$, $M v = \Omega$. 
Note that positions $\vec{r}$ with Latin/Greek indices denote coordinates of
defects/mesh-points. The discretized probability function then derives from the
measure Eq.~\eqref{eq:DUr},
\begin{align}\label{eq:PUa}
   \mathcal{P}[\{U_{\alpha}\}] &= 
   \int \Big[\prod_{i=1}^{N} \frac{d^{2} r_{i}}{\Omega} \Big] 
   \Big\{\prod_{\beta} \delta[U_\beta - U(\vec{r}_{\beta})]\Big\}.
\end{align}
We rewrite the Dirac $\delta$ distributions in Fourier space and obtain the expression
\begin{align}\nonumber
   \mathcal{P}[\{U_{\alpha}\}] 
   &= \int \!\Big[\!\prod_{\alpha} \frac{dK_{\alpha}}{2\pi/v} \Big] 
   \Big[\prod_{i}^{N} \frac{d^{2} r_{i}}{\Omega} \Big] 
   e^{i v \sum_{\beta} \!K_{\beta} [U_\beta - U(\vec{r}_{\beta})]}\\ \nonumber
   &= \int \Big[\prod_{\alpha} \frac{dK_{\alpha}}{2\pi/v}\Big] 
   e^{i v \sum_{\beta} K_{\beta} U_{\beta}}
   \\[-0.3em]
   &\qquad\quad \times \bigg[ \int \frac{d^{2} r}{\Omega} 
   e^{-i v \sum_{\beta} K_{\beta} V(\vec{r}_{\beta} - \vec{r})} \bigg]^{N},
\end{align}
where we have made use of Eq.~\eqref{eq:Ur-app}.  Adding and subtracting unity in
the last square bracket, and taking the thermodynamic limit $N, \Omega \!\to\!
\infty$ with $n_{p} = N/\Omega$, we can rewrite the above equation as
\begin{align}
   \!\!\!\!
   \mathcal{P}[\{U_{\alpha}\}] \!&=\!\! 
   \int \!\Big[\!\prod_{\alpha} \frac{dK_{\alpha}}{2\pi/v}\Big]
   e^{\psi[\{K_{\alpha}, U_{\alpha}\}; V(\vec{r})]}
\end{align}
with
\begin{align}\label{eq:prob-distr-to-expand}
   \psi[\{K_{\alpha}, U_{\alpha}\}; V(\vec{r})] &= 
   i v \sum_{\beta}\! K_{\beta} U_{\beta} \\[-0.3em] \nonumber
   &\quad+ n_{p} v \sum_{\alpha} \Big[ e^{-i v \sum_{\beta}\! K_{\beta} 
   V(\vec{r}_{\beta} - \vec{r}_{\alpha})} - 1 \Big].
\end{align}
For consistency, we have discretized the average over defect positions
$\int d^{2} r \to v \sum_{\alpha}$.
The saddle-point equation $\partial \psi / \partial K_{\beta} = 0$ for a given
$K_{\beta}$ reads
\begin{align}\label{eq:saddle-point-equation}
   U_{\beta} = n_{p} v \sum\nolimits_{\alpha} V(\vec{r}_{\beta} - \vec{r}_{\alpha}) 
   e^{-i v \sum_{\gamma}\! K_{\gamma} V(\vec{r}_{\gamma} - \vec{r}_{\alpha})}.
\end{align}
We expand the exponential function above assuming its argument to be small, an
assumption that will be validated a-posteriori below, and find
\begin{align}\label{eq:saddle-point-equation-expanded}
   U_{\beta} &\approx n_{p} v \sum_{\alpha} V(\vec{r}_{\beta} \!-\! \vec{r}_{\alpha})\\[-0.2em] 
   \nonumber
   &\quad -\! i n_{p} v^{2} \sum_{\alpha,\gamma} K_{\gamma} 
   V(\vec{r}_{\beta} \!-\! \vec{r}_{\alpha}) 
   V(\vec{r}_{\gamma} \!-\! \vec{r}_{\alpha}) \\[-0.2em] \nonumber
   &\quad -\! n_{p} v^{3}\! \sum_{\alpha,\gamma,\delta} \! K_{\gamma} K_{\delta} 
   V(\vec{r}_{\beta} \!-\! \vec{r}_{\alpha})  V(\vec{r}_{\gamma} \!-\! \vec{r}_{\alpha}) 
   V(\vec{r}_{\delta} \!-\! \vec{r}_{\alpha}).
\end{align}
The first term on the right-hand side is the potential's mean value which we have
assumed to vanish. For the second term in the
expression above, we introduce
\begin{align}\label{eq:Green_disc}
   G_{\beta,\gamma} = n_{p} v \sum\nolimits_{\alpha} V(\vec{r}_{\beta} - \vec{r}_{\alpha}) 
   V(\vec{r}_{\gamma} - \vec{r}_{\alpha}),
\end{align}
the discrete version of the two-point correlator \eqref{eq:Green}. With
$G_{\beta,\gamma}$ of the scale $(n_{p}\xi^{2}) V_{0}^{2}$ and decaying on a
length $|\vec{r}_{\beta} - \vec{r}_{\gamma}| \sim \xi$, we arrive at the
estimate
\begin{align}\label{eq:K}
   \bar{K}_{\beta} \equiv v\!\!\!\!\!\!\sum_{\gamma, |\vec{r}_{\beta} - \vec{r}_{\gamma}| 
   {\scriptscriptstyle <} \xi\!\!}  \!\!\!\!\!\! K_{\gamma}
   \sim \frac{U_{\beta}}{(n_{p}\xi^{2}) V_{0}^{2}}.
\end{align}
Substituting this estimate in the third term of Eq.~\eqref{eq:saddle-point-equation-expanded}, we find that it is small when
\begin{align}\label{eq:VK}
   V_{0} \bar{K}_{\beta} \ll 1.
\end{align}
The width of the distribution function for the expectation value of the
potential grows only with $(n_{p}\xi^{2})^{1/2} V_{0}$, what tells us that in
the limit $n_{p}\xi^{2} \to \infty$, the above condition is satisfies almost
everywhere (except for far-distant tails: for $U \sim V_0 n_p \xi^2$, see
\eqref{eq:K} and \eqref{eq:VK}, the probability has dropped to $\exp(-U^2/G)
\sim \exp(- \mathrm{const.} \,n_p \xi^2)$). This reasoning justifies the
truncation of \eqref{eq:saddle-point-equation-expanded} to include only terms
up to linear order in $K$. At the same time, it validates the assumption used
after Eq.~\eqref{eq:saddle-point-equation} and allows to expand the
exponential in \eqref{eq:prob-distr-to-expand} to quadratic order in $K$. We
thus arrive at the simple expression
\begin{align}
   \mathcal{P}[\{U_{\alpha}\}] \!&\approx\!\! \int \!\Big[\!\prod_{\alpha} 
   \frac{dK_{\alpha}}{2\pi/v}\Big] e^{i v \sum_{\beta}\! K_{\beta} U_{\beta}}
      e^{- \frac{1}{2} v^{2} \sum_{\beta, \gamma} K_{\beta} G_{\beta, \gamma} K_{\gamma}}
\end{align}
for the discretized probability distribution, a result that becomes exact for
$n_{p} \xi^{2} \!\to\! \infty$. Computing the Gaussian integrals over $K$, we
find that
\begin{align}\label{}
   \mathcal{P}[\{U_{\alpha}\}] &\propto
   e^{- \frac{1}{2} \sum_{\beta, \gamma} 
   U_{\beta} (G^{-1})_{\beta, \gamma} U_{\gamma}},
\end{align}
where we have used the discrete version of the inversion identity $\int d^{2}x \, G(\vec{r}-
\vec{x})\, G^{-1}(\vec{r}'-\vec{x}) = \delta(\vec{r}-\vec{r}')$, i.e.,
\begin{align}\label{eq:inversion}
   v \sum\nolimits_\beta G_{\alpha, \beta}(G^{-1})_{\beta, \gamma} = \delta_{\alpha, \gamma}/v.
\end{align}
Returning back to the continuum notation, we arrive at the final result
\begin{align}
   \mathcal{P}[U(\vec{r})] \to \mathcal{P}_{\mathrm{G}}[U(\vec{r})] 
   \equiv \frac{1}{\mathcal{Z}} e^{-\frac{1}{2} \int \frac{d^{2} r}{\Omega} 
   \frac{d^{2} r'}{\Omega} U(\vec{r}) G^{-1}(\vec{r} - \vec{r}') U(\vec{r}')},
\end{align}
where $\mathcal{Z}$ accounts for the correct normalization.
Note that more terms in the
expansion of \eqref{eq:prob-distr-to-expand} need to be retained if one is
interested in properties away from the body of the probability distribution,
at least in principle.

\section{Parabolic traps}\label{app:parabola}

The determination of the probability distribution $p(\D,\T)$ of the Hessian
determinant and trace for a landscape made from cut parabolas makes use of
Eqs.~\eqref{eq:pi-ABC} and \eqref{eq:DUr}, from which follows that
\begin{align}\label{eq:ABC-probability-distribution}
   \pi(a,b,c) = \int & \Big[\prod_{j=1}^{N} \frac{d^{2} r_{j}}{\Omega}\Big]
   \, \delta[U_{xx}(0) - 2a] \\[-0.2em]\nonumber
   &\qquad \times \delta[U_{yy}(0) - 2b]\, \delta[U_{xy}(0) - c].
\end{align}
Rewriting the delta-distributions in Fourier space, and expressing the
potential $U(\vec{r})$ through the sum of individual defect potentials
$V(\vec{r}-\vec{r}_i)$, we obtain the expression
\begin{align}\label{eq:fourier-expression_0}
   \pi(a,b,c) &= \int \frac{dk\, dl\, dm}{(2\pi)^{3}} \,e^{i(2ka + 2lb + mc)}
   \\[-0.2em] \nonumber
   &\times \Big[ \int \frac{d^{2}r}{\Omega}
   e^{-i [k V_{xx}(\vec{r}) + l V_{yy}(\vec{r}) + m V_{xy}(\vec{r})]}\Big]^N.
\end{align}
In the thermodynamic limit, $N, \Omega \!\to\! \infty$ at fixed defect density
$n_{p} \!=\! N/\Omega$, the last factor can be rewritten as
\begin{align}\label{eq:exp_epsilon}
   \Big[1+ \frac{n_p}{N} \epsilon(k,l,m)\Big]^N = e^{n_p \epsilon(k,l,m)}
\end{align}
with $\epsilon(k,l,m)$ involving only the potential shape $V(\vec{r})$ of an
individual defect, 
\begin{align}\label{eq:epsilon}
   \epsilon(k,l,m) = \! \int \!\! d^{2} r \big[ e^{-i [k V_{xx}(\vec{r}) 
   + l V_{yy}(\vec{r}) + m V_{xy}(\vec{r})]}\! - \! 1\big].
\end{align}
As a result, we arrive at the compact form
\begin{align}\label{eq:pi-most-general}
   \pi(a,b,c) = \! \int  \!\frac{dk\, dl\, dm}{(2\pi)^{3}} e^{i(2ka + 2lb + mc)} 
   e^{n_{p} \epsilon(k,l,m)}.
\end{align}
While the above procedure applies for all defect types, we explicitly evaluate the
above expressions for the cut parabolic defect. Since all second derivatives of
$V(\vec{r})$ are either $2 V_0/\xi^2$ or zero, Eq.~\eqref{eq:epsilon}
reads
\begin{align}\label{eq:epsilon-harm}
   \epsilon(k,l,m) = \Omega_{0} \big[ e^{-i2V_0(k + l)/\xi^2}  - 1\big].
\end{align}
Inserting this result into Eq.~\eqref{eq:pi-most-general}, expanding the
last factor in a power series, and using the binomial theorem, we find that
\begin{align}\label{}
   \pi(a,b,c) &= \int \frac{dk\, dl\, dm}{(2\pi)^{3}} \,
       e^{i(2ka + 2lb + mc)}\\[-0.2em]\nonumber
       &\quad \sum_{\nu=0}^{\infty} \sum_{\mu=\nu}^{\infty} 
       \frac{(n_{p}\Omega_{0})^{\mu}}{\nu! (\mu-\nu)!}
       e^{-i 2V_0 \nu (k + l)/\xi^2} (-1)^{\mu-\nu}.
\end{align}
The integrations over $k,l,m$ provide $\delta$ distributions and rearranging terms
in the  sum, we obtain
\begin{align}\label{eq:pi_par}
   \pi(a,b,c) &\!=\! \sum\nolimits_{\nu=0}^{\infty} \delta[ \nu (2V_0/\xi^2) - 2a] \, 
    \delta[ \nu (2V_0/\xi^2) - 2b] \nonumber \\[-.3em]
            & \hspace{4em}
            \times \delta(c) \,
       \Poiss(\nu, n_{p}\Omega_{0}).
\end{align}
As expected, the Hessian matrix can only take on discrete values $(2\nu
V_0/\xi^2)\, \mathbb{I}$ and, correspondingly, the probability distribution is a
sum of $\delta$ distributions. 

Next, we make use of the result for $\pi(a,b,c)$, Eq.~\eqref{eq:pi_par}, in
the determination of the probability distribution $p(\D,\T)$ for a Hessian $H$
with $\det H = \D$ and $\tr H = \T$, see Eq.~\eqref{eq:pXY}. The
expression \eqref{eq:X-prob-ABC} for $p(\D)$ generalizes to
\begin{align}\label{eq:pXY-pi}
   p(\D,\T) \! = \!\!\! \iiint\limits_{\!\text{-}\infty\,\text{-}
   \infty\,\text{-}\infty}^{\quad\infty\infty\infty}
   \!\!\! da\, db\, dc\ \pi(a,b,c) f(a,b,c;\D,\T)
\end{align}
with
\begin{align}\label{eq:fABCXY}
   f(a,b,c;\D,\T) = \delta[4ab\!-\!c^{2} \!-\! \D]\, \delta[ 2a\! +\! 2b\! -\! \T].
\end{align}
and inserting the result Eq.~\eqref{eq:pi_par} for $\pi(a,b,c)$, we find
\begin{align}
   p(\D,\T) &= \sum\nolimits_{\nu=0}^{\infty} \Poiss(\nu, n_{p}\Omega_{0})
        \delta[\nu^{2} (2V_0/\xi^2)^{2} - \D]  \nonumber \\[-0.3em]
        \label{eq:p_par}
        & \hspace{4.2em}
        \times \delta[\nu (2V_0/\xi^2) - \T].\\[-1em] \nonumber
\end{align}
The final integration over strictly positive $\D$ and $\T$ results in the
stable area fraction of the Hessian map, $\mathcal{C}_{\mathrm{pos}} = 1 -
\Poiss(0,n_{p}\Omega_{0})$.

\vfill
\pagebreak
\twocolumngrid
\bibliographystyle{apsrev4-1-titles}
\bibliography{maze-Hessian}

\end{document}